\documentclass[12pt,preprint,numberedappendix,twocolappendix]{emulateapj}

\newcommand\bibinc{n}		

\usepackage{subeqnarray}
\usepackage{amsmath}
\usepackage{hyperref}
\usepackage{ulem}
\usepackage{stmaryrd}

\hypersetup{colorlinks=true,linkcolor=black,citecolor=blue,urlcolor=black}

\usepackage{natbib}
\def\tt{\bf   }

\newcommand{\Sec}[1]{Section~\ref{#1}}

\newcommand{\Fig}[1]{Figure~\ref{#1}}

\setcitestyle{citesep={,}}
\begin{document}

\slugcomment{Submitted to AAS Journals}

\shorttitle{Ultra-hot Jupiter Atmospheres}
\shortauthors{Tan \& Komacek}

\title{The Atmospheric Circulation of Ultra-Hot Jupiters}
\author{Xianyu Tan$^{1}$ and Thaddeus D. Komacek$^{2}$} \affil{ $^1$Atmospheric, Ocean, and Planetary Physics, Department of Physics, Oxford University, OX1 3PU, United Kingdom \\ 
$^2$Department of the Geophysical Sciences, The University of Chicago, Chicago, IL, 60637 \\ 
\url{xianyu.tan@physics.ox.ac.uk}}
\begin{abstract}
Recent observations of ultra-hot Jupiters with dayside temperatures in excess of $2500 \mathrm{K}$ have found evidence for new physical processes at play in their atmospheres. In this work, we investigate the effects of the dissociation of molecular hydrogen and recombination of atomic hydrogen on the atmospheric circulation of ultra-hot Jupiters. To do so, we incorporate these effects into a general circulation model (GCM) for hot Jupiter atmospheres, and run a large suite of models varying the incident stellar flux, rotation period, and strength of frictional drag. We find that including hydrogen dissociation and recombination reduces the fractional day-to-night temperature contrast of ultra-hot Jupiter atmospheres and causes the speed of the equatorial jet to decrease in simulations with fixed rotation. This is because the large energy input required for hydrogen dissociation cools the dayside of the planet, and the energy released due to hydrogen recombination warms the nightside. The resulting decrease in the day-to-night temperature contrast reduces the day-to-night pressure gradient that drives the circulation, resulting in weaker wind speeds. The results from our GCM experiments qualitatively agree with previous theory which found that the fractional day-night temperature contrast of ultra-hot Jupiters should decrease with increasing equilibrium temperature due to hydrogen dissociation and recombination. Lastly, we compute full-phase light curves from our suite of GCM experiments, finding that the reduced day-to-night temperature contrast in ultra-hot Jupiter atmospheres causes a smaller phase curve amplitude. The reduction in phase curve amplitude due to hydrogen dissociation and recombination could explain the relatively small phase curve amplitudes of observed ultra-hot Jupiters.
\end{abstract}
\section{Introduction}
\label{sec:introvar}
\indent Ultra-hot Jupiters, gas giant exoplanets with equilibrium temperatures\footnote{The equilibrium temperature here is defined as the temperature of a spherical blackbody planet that fully redistributes heat received from the star over the globe, which can be expressed as $T_{\rm{eq}}=[F_{\star}/(4\sigma)]^{1/4}$, where $F_{\star}$ is the total incoming stellar flux and $\sigma$ is the Stefan-Boltzmann constant.}  in excess of $2200~\mathrm{K}$, represent a new regime of exoplanets to observationally characterize and understand. Recent observations of ultra-hot Jupiters \citep{Stevenson:2014aa,Haynes:2015,Beatty:2017aa,Evans:2017aa,Zhang:2017a,Arcangeli:2018aa,Kreidberg:2018aa,Mansfield:2018aa,Arcangeli:2019aa,Bell:2019aa} have found that their emission spectra  differ from those of normal hot Jupiters. The emission spectra of ultra-hot Jupiters are nearly featureless, because water dissociates on the daysides of ultra-hot Jupiters \citep{Kitzmann:2018aa,Kreidberg:2018aa,Lothringer:2018aa,Parmentier:2018aa}. In addition, molecular hydrogen (H$_2$) begins to partially dissociate at temperatures $\gtrsim 2500~\mathrm{K}$, and as a result hydrogen is partially dissociated in the atmospheres of ultra-hot Jupiters.    \\
\indent The dissociation of hydrogen requires energy input from the atmosphere to break the strong H$_2$ bond. After atomic hydrogen is transported by atmospheric dynamics to cooler regions of the atmosphere, the recombination from atomic to molecular hydrogen releases a significant amount of heat. This heat release is approximately a hundred times greater  than water's latent heat of condensation \citep{Bell:2018aa}. The impact of hydrogen dissociation and recombination on the atmospheric temperature structure of ultra-hot Jupiters was first studied by \cite{Bell:2018aa}. To do so, \citeauthor{Bell:2018aa} used a semi-analytic energy transport model similar to that of \cite{Cowan:2011} but including hydrogen dissociation and recombination as an additional energy flux term. \citeauthor{Bell:2018aa} found that the heating from hydrogen recombination could significantly warm the nightside of ultra-hot Jupiters, leading to reduced phase curve amplitudes and increased phase curve offsets. Additionally, \cite{Komacek:2018aa} incorporated hydrogen dissociation and recombination into the analytic model of \cite{Komacek:2015,Zhang:2016}, and \cite{Komacek:2017} for the phase curve amplitudes of hot Jupiters. \citeauthor{Komacek:2018aa} found that the reduced phase curve amplitude due to hydrogen dissociation and recombination may explain the relatively low phase curve amplitudes from \textit{Hubble} and \textit{Spitzer} observations of WASP-103b \citep{Kreidberg:2018aa} and WASP-33b \citep{Zhang:2017a}. \\
\indent Both \cite{Bell:2018aa} and \cite{Komacek:2018aa} studied heat transport using a one-dimensional framework. There has been a wide range of previous work studying the three-dimensional (3D) effects of atmospheric circulation on the resulting climate of hot Jupiters \citep{showman_2002,Cooper:2005,Menou:2009,Showmanetal_2009,Showman_2009,Rauscher:2010,Thrastarson:2010,Heng:2011,Showman_Polvani_2011,perna_2012,Rauscher_2012,Dobbs-Dixon:2013,showman_2013_doppler,Kataria:2014,Mayne:2014,Fromang:2016,Kataria2016,Mayne:2017,Drummond:2018,Koll:2017,Mendonca:2018,Steinrueck:2018aa}. However, the effects of hydrogen dissociation and recombination on the 3D atmospheric dynamics of ultra-hot Jupiters has not been studied to date.  \\
\indent In this work, we incorporate hydrogen dissociation and recombination into an established GCM to simulate how the atmospheric circulation of ultra-hot Jupiters differs from normal hot Jupiters. \Sec{sec:methods} describes how we incorporate the effects of hydrogen dissociation and recombination into the GCM, along with our numerical setup and parameter space exploration. We describe our GCM results for how the atmospheric circulation and simulated phase curves are affected by hydrogen dissociation and recombination in \Sec{sec:results}. To summarize, we find that hydrogen dissociation causes the dayside to cool and recombination causes the nightside to warm. This significantly reduces the phase curve amplitude, as expected by \cite{Bell:2018aa} and \cite{Komacek:2018aa}. We compare our simulated phase curves to observations in \Sec{sec:comptoobs}. Lastly, we discuss our results in \Sec{sec:disc} and state key takeaways from this work in \Sec{sec:conc}. 
\section{Model} \label{sec:methods}


We have developed an idealized GCM  to incorporate the dynamical effects of hydrogen dissociation and recombination, including the cooling and heating released from the chemical reaction and the effects of dissociation and recombination on the atmospheric mean molecular weight and specific heat. The major difference of our model compared to previous hot Jupiter GCMs is that we include more generalized thermodynamics and the significant change of mean molecular weight when two dominant gas species are comparable in mass. We exclude the effects of varying opacity due to the changing abundance of H$^-$ \citep{Parmentier:2018aa, Lothringer:2018aa} and due to the presence of clouds (e.g., \citealp{Lee:2016,Parmentier:2015,Lines:2018,Roman:2019aa}), and we do not include magnetohydrodynamic effects \citep{Perna_2010_1,Menou:2012fu,batygin_2013,Rauscher_2013,Rogers:2014,Rogers:2017,Hindle:2019aa} in this study. We use a simplified modeling approach to focus on improving the understanding of fundamental processes in complex exoplanetary atmospheres, which differs from the highly comprehensive models used for making precise predictions.

\subsection{Dynamics}
The dynamical core of the GCM solves the global, three-dimensional hydrostatic primitive equations that govern the large-scale flow in stratified atmospheres, including  the horizontal momentum, hydrostatic equilibrium and continuity equations in pressure coordinates as follows, respectively,
\begin{equation}
\frac{d\mathbf{v}}{dt} = - f\hat{k} \times \mathbf{v} - \nabla_p \Phi + \mathcal{D}_{\mathbf{v}}, 
\label{eq.momentum}
\end{equation}
\begin{equation}
\frac{\partial \Phi}{\partial p} = -\frac{1}{\bar{\rho}}, 
\label{eq.hydrostatic}
\end{equation}
\begin{equation}
\nabla_p \cdot \mathbf{v} + \frac{\partial\omega}{\partial p} = 0,
\label{eq.cont}
\end{equation}
where $p$ is pressure, $\mathbf{v}$ is the horizontal velocity vector on isobars, $\omega=dp/dt$ is the vertical velocity in pressure coordinates, $f=2\Omega\sin \phi$ is the Coriolis parameter (here $\phi$ is latitude and $\Omega$ is the planetary rotation rate), $\Phi$ is the geopotential, $\hat{k}$ is the local unit vector in the vertical direction, $\bar{\rho}$ is the mean gas density, $\nabla_p$ is the horizontal gradient in pressure coordinates, and $d/dt=\partial/\partial t + \mathbf{v}\cdot\nabla_p + \omega\partial/\partial p$ is the material derivative. A thermodynamics energy equation is needed to close the system and will be separately introduced below. 

We include a frictional drag term  $\mathcal{D}_{\mathbf{v}}$ to the horizontal momentum equations, the same as that in \cite{Komacek:2015}. This drag scheme includes two parts, one of which is a basal drag that exists from the bottom boundary to 10 bars, representing momentum mixing with the relatively quiescent interior. The other part is a spatially independent drag which is characterized by a spatially independent drag timescale $\tau_{\rm{drag}}$, mimicking the effects of magnetohydrodynamics and turbulence. The dissipated kinetic energy transfers to thermal energy and is added to the thermodynamics energy equation. In this study we vary $\tau_{\rm{drag}}$ to investigate the circulation under different drag regimes.
 
\subsection{Fractional atomic hydrogen}
We use a scalar tracer $q=\rho_H/(\rho_H + \rho_{H_2})$ to represent the mass mixing ratio of atomic hydrogen H relative to the total gas (H + $\rm{H}_2$). The following tracer equation tracks the global evolution of the mass mixing ratio of atomic hydrogen, which is solved in the dynamical core of our GCM:
\begin{equation}
\frac{dq}{dt} = -\frac{\delta q}{\tau_{\rm{relax}}},
\label{eq.tracer}
\end{equation}
where $\tau_{\rm{relax}}$ is a relaxation timescale.
The change of $q$ feeds back into dynamics via the release and absorption of heat, the change of the atmospheric mean molecular weight, and the change of the mean atmospheric heat capacity. The change of the mass fraction of atomic hydrogen due to a change of temperature or pressure $\delta q$ is expressed as 
\begin{equation}
\delta q= \frac{q-q_{\rm{eq}}}{1+ \frac{\mathcal{L}_h}{\bar{c_p}} \frac{\partial q_{\rm{eq}}}{\partial T}},
\end{equation}
where $T$ is temperature, $q_{\rm{eq}}$ is the equilibrium H mass fraction set to be a function of pressure and temperature only, $\mathcal{L}_h$ is the heat per unit mass released from hydrogen recombination and $\bar{c_p}$ is the mass-weighted mean specific heat. An extra term in the denominator comes from the fact that the heat release/absorption changes the air temperature and prevents a full relaxation to $q_{\rm{eq}}$ at the original temperature. 
For numerical stability, the change of $q$ is not taken to be the full amount of $\delta q$ in a dynamical time step, but instead is relaxed towards  $\delta q$ over a relaxation timescale $\tau_{\rm{relax}}$, which is longer than a dynamical time step. The chemical timescale of $\rm{H}_2$ thermal dissociation is expected to be much shorter than the dynamical timescale in atmospheres of ultra-hot Jupiters, and the chemistry is expected to be nearly in equilibrium especially near the photosphere\footnote{For example, at $10^{-3}$ bar, the chemical timescale of $\rm{H}_2$ thermal dissociation is between $10^2$ to $10^3$ s at 2000 K, and between $10$ to $10^2$ s at 3000 K. At $10^{-1}$ bar, the timescale is well below 1 s for temperature higher than 1800 K (\citealp{tsai2018}, Shang-Min Tsai, personal communication).}. As a result, in our simulations $\tau_{\rm{relax}}$ is set to be 1.5 times the dynamical time to satisfy both constraints.

Considering thermal dissociation alone, the equilibrium mixing ratio of atomic hydrogen gas $q_{\rm{eq}}$ can be derived from the Saha equation (e.g., as in Appendix A of \citealp{berardo2017}). The molar fraction of atomic hydrogen relative to the total gas is denoted $\chi_H$, which is given by 
\begin{equation}
\chi_H = \sqrt{(1-\chi_H) \hat{Y}},
\label{eq.chi1}
\end{equation}
where 
\begin{equation}
\hat{Y}=\frac{2\Theta_{rot}}{p}k_b^{5/2}\frac{(\pi m_H T)^{3/2}}{h^3}e^{-\epsilon/(k_bT)}.
\label{eq.y}
\end{equation}
In Equation (\ref{eq.y}), $\Theta_{rot}=85.4~K$, $k_b$ is the Boltzmann constant, $m_H$ is the mass of a hydrogen atom, $h$ is the Planck constant and $\epsilon=7.148\times 10^{-19} ~J$ is the hydrogen dissociation energy in Joules (this implies that $\mathcal{L}_h$ is taken to be $2.18\times10^8 \ \mathrm{J} \ \mathrm{kg}^{-1}$). Equation (\ref{eq.chi1}) can be written as 
\begin{equation}
\chi_H = \frac{1}{2} (\sqrt{\hat{Y}}\sqrt{\hat{Y}+4} - \hat{Y}),
\end{equation}
which is equivalent to Equation (3) of \cite{Bell:2018aa}\footnote{Note that our $\hat{Y}$ is the inverse of $Y$ from \cite{Bell:2018aa}.}. The equilibrium mass fraction of atomic hydrogen is simply $q_{\rm{eq}}=\chi_H/(2-\chi_H)$.

\subsection{Thermodynamics}
We use a modified potential temperature $\theta'$, a thermodynamic quantity that is conserved in adiabatic flow in condensible atmospheres, as a prognostic thermodynamic variable instead of the standard potential temperature $\theta = T(p_s/p)^{\kappa}$ used in a dry, non-condensible atmosphere, where $\kappa=R/c_p$ is the ratio between the specific gas constant and the specific heat for dry air and $p_s=1 ~\rm{bar}$ is a reference pressure. Similar to \cite{pierrehumbert2016}, the conserved quantity in adiabatic flow can be derived from the first law of thermodynamics for a mixture of two ideal-gas components (which are considered to be H and $\rm{H_2}$ gas):
\begin{equation}
\delta Q = \bar{c_p} dT-\frac{1}{\bar{\rho}} dp =  \bar{c_p} dT-\frac{\bar{R}T}{p} dp.
\label{eq.firstlaw}
\end{equation}
 The ideal gas law  $p=\bar{\rho} \bar{R} T$ is assumed for the equation of state for the atmosphere where $\bar{R}$ is the mean specific gas constant, which together with $\bar{c_p}$ are weighted by fractional mass as defined below:
\begin{equation}
\bar{R} = R_{H_2}(1-q)+R_H q, 
\label{eq.R}
\end{equation}
and \begin{equation}
\bar{c_p} = c_{p,H_2}(1-q)+c_{p,H}q, 
\label{eq.c_p}
\end{equation}
where $c_{p,H_2}$ and $c_{p,H}$ are the specific heat at constant pressure and $R_{H_2}$ and $R_{H}$ are the specific gas constants for molecular and atomic hydrogen gas, respectively. We assume a fixed specific heat for both species, with $c_{p,H_2}=13000 \ \mathrm{J} \ \mathrm{kg}^{-1} \mathrm{K}^{-1}$ and $c_{p,H}=20790 \ \mathrm{J} \ \mathrm{kg}^{-1} \mathrm{K}^{-1}$, and we assume $\kappa=2/7$ for $\rm{H_2}$ gas. In reality the specific heat of molecular hydrogen slightly increases with increasing temperature, but it is assumed to be a constant in this study. We do so because we want to restrict our problem to explore only the effects of change in the relative H and $\rm{H_2}$ gas fractions. Combining Equations (\ref{eq.firstlaw}), (\ref{eq.R}), and (\ref{eq.c_p}), we find that $\theta'=T(p_s/p)^{\kappa'}$ is conserved in adiabatic flow ($\delta Q=0$), where
\begin{equation}
\kappa'=\frac{R_{H_2}(1-q)+R_H q}{c_{p,H_2}(1-q)+c_{p,H}q}.
\end{equation}
The modified potential temperature $\theta'$ is then the prognostic variable of the thermodynamics equation solved in the GCM. 

The diabatic heating/cooling rate $dQ/dt$ includes contributions from the chemical heat release of H-$\rm{H}_2$ conversion\footnote{An interesting consequence of the chemical heat release from hydrogen recombination is the decrease of the temperature lapse rate $d\ln T/d\ln p$ along the parcel trajectory without heat exchange with the environment. This is an analog to moist processes in condensible atmospheres, and its impact could be  significant in the hot convective interior because the chemical heat release from hydrogen recombination is a few orders of magnitude larger than that of latent heat release from water phase change. The fractional difference between the temperature lapse rate accounted for by hydrogen dissociation and the adiabatic lapse rate of a dry molecular hydrogen atmosphere ($d\ln T/d\ln p = \kappa=2/7$) can reach up to several tens of percent depending on the dissociation fraction. This may affect convection in convective atmospheres like those of brown dwarfs and young giant planets, but is likely not relevant near the photospheres of ultra-hot Jupiters which are likely strongly stratified.} ($\mathcal{L}_h \delta q/\tau_{\rm{relax}}$), the radiative heating, and dissipated kinetic energy from frictional drag.  
The strong day-night radiative forcing is the main driver of the global circulation of hot Jupiters through diabatic heating. Here we utilize a semi-grey radiative transfer scheme to model the radiative forcing. The detailed implementation is discussed in Section 3 of \cite{Komacek:2017}. In this study the atmosphere is considered absorptive only and we neglect scattering of radiation due to molecules and cloud particles. The opacity profiles for both the visible and thermal band are prescribed as a function of pressure only, and remain the same throughout all simulations in this study. The thermal opacity profile as a function of pressure is 
\begin{equation}
\log_{10}\kappa_{\rm{th}} = 0.0498(\log_{10}p)^2 - 0.1329\log_{10}p - 2.9457,
\end{equation}
 and the visible opacity profile is 
 \begin{equation}
 \log_{10} \kappa_{\rm{v}} = 0.0478(\log_{10}p)^2 - 0.1366\log_{10}p - 3.2095,
 \end{equation}
 in which the opacity has a unit of $\rm{m^2kg^{-1}}$ and pressure is in units of Pa. The functional form of the opacity comes from a quadratic fit in $\log \kappa$-$\log p$ space to the Rosseland-mean opacity profile from a self-consistently generated radiative-equilibrium T-P profile with $T_{\rm{eq}}=2000$ K and $T_{\rm{int}}=200$ K. In generating this equilibrium T-P profile, we couple the Rosseland-mean opacity tables from \cite{Freedman:2014} to the radiative transfer, and then integrate the profile forward with time until a radiative equilibrium is reached. The thermal opacity slightly exceeds the visible opacity at all pressures, resulting in a ``typical'' hot-Jupiter thermal profile that has increasing temperature with increasing pressure at most pressure levels. Thermal dissociation of molecules can change the opacity substantially. In this study we intentionally fix the opacity profiles, which can give rise to similar temperature lapse rates $d\ln T/d\ln P$ with different planetary equilibrium temperatures. Because the temperature lapse rate is one of the key factors controlling atmospheric dynamics, this is a better strategy to isolate the effects of varying stellar irradiation and atomic/molecular hydrogen fractions than allowing a self-consistent coupling of opacity to temperature or composition. \\
 \indent Having solved the total radiative flux $F$, the radiative heating/cooling rate is then given by $\frac{g}{\bar{c_p}}\frac{\partial F}{\partial p}$ where $g$ is the surface gravity.
Finally, accounting for all the heating/cooling terms, the thermodynamics equation solved in the GCM is
\begin{equation}
\frac{d\theta'}{dt} = \left(\frac{p_s}{p} \right)^{\kappa'} \frac{1}{\bar{c_p}} \left(\mathcal{L}_h\frac{\delta q}{\tau_{\rm{relax}}} + g\frac{\partial F}{\partial p}  + \mathcal{D}_{T} \right) ,
\label{eq.thermo}
\end{equation}
in which $\mathcal{D}_{T}$ represents the dissipated kinetic energy by the drag in equation (\ref{eq.momentum}).

\subsection{Mean molecular weight}
The mean molecular weight of air  can decrease substantially when the  fractional atomic hydrogen becomes large. This affects large-scale dynamics via an increase of the atmospheric scale height in areas with large atomic hydrogen fractions, which then affect the horizontal geopotential gradient. This effect is implemented in the dynamical system through the change of mean gas density  $\bar{\rho}$ in the hydrostatic equation (\ref{eq.hydrostatic}) as follows: 
\begin{equation}
\begin{aligned}
\delta \Phi &= -\frac{1}{\bar{\rho}}\delta p = -\frac{\bar{R}T}{p}\delta p = -\bar{R} \theta' \left(\frac{p}{p_s}\right)^{\kappa'} \frac{\delta p}{p} \\
& = -R_{H_2}(1+q(\frac{R_ H}{R_{H_2}}-1)) \theta' p_s^{-\kappa'} \frac{1}{\kappa'} \delta(p^{\kappa'}) \\
& = -c_{p,H_2}(1+q(\frac{c_{p, H}}{c_{p,H_2}}-1)) \theta' \delta\left(\frac{p}{p_s}\right)^{\kappa'}  .
\end{aligned}
\end{equation}
Given the large day-night difference in the atomic hydrogen fraction of ultra-hot Jupiter atmospheres, the day-night geopotential difference increases due to the change in atmospheric mean molecular weight from dayside to nightside. Compared to models that do not include the mean molecular weight effect but with the same day-night radiative forcing, models including the change in mean molecular weight from dayside to nightside have increased wind speeds and  thus could further reduce day-night temperature differences.

\subsection{Numerical details}
\label{sec.numerical}
We focus on understanding the atmospheric circulation in a general sense instead of on specific targets, therefore the planetary parameters are assumed to be similar to that of typical hot Jupiters. We adopt a planetary radius of $1.05\times 10^8$ m and a surface gravity of $g=11 ~ \rm{ms^{-2}}$ for all of our simulations. In the main suite of simulations, we also fix the planetary rotation rate (set equal to the orbital period) to be 2.43 days, equivalent to that of a planet with an equilibrium temperature of $T_{\rm{eq}} \sim 2000$ K orbiting a star with an effective temperature of 6000 K, mass of 1.2 times that of the Sun, and radius of 1.8 times that of the Sun, to represent a typical F star. We perform two types of models that are with or without the dynamical effects of hydrogen dissociation and recombination included but with all other parameters the same. In each type of models we perform two sets of models with either  weak drag ($\tau_{\rm{drag}}=10^7$ s) or  strong drag ($\tau_{\rm{drag}}=10^4$ s)\footnote{We define ``strong'' drag as when the drag timescale is comparable to or shorter than other relevant dynamical timescales, including the rotation period and advective timescale.}. Comparison between these two types of models with a fixed rotation period allows us to understand the dynamical effects of hydrogen dissociation alone.  We systematically vary the equilibrium temperature of our simulated planets from 1600 K to 3600 K to determine the dependence of circulation on incident stellar flux. We set an internal temperature of $T_{\rm{int}}=200$ K, which results in an internal heat flux at the bottom of the domain given by $F_{\rm{int}}=\sigma T_{\rm{int}}^4$.  In reality, some ultra-hot Jupiters rotate substantially faster than the canonical hot Jupiters because they are closer to their host stars. To investigate the additional dynamical effects of the increasing rotation rate on the atmospheric circulation,  we also present a set of models with different $T_{\rm{eq}}$ but with a self-consistently decreasing rotation period with increasing equilibrium temperature.
 
The dynamical equations are solved in a global cubed-sphere grid using an atmospheric general circulation model, the MITgcm (\citealp{Adcroft:2004}, see also \url{mitgcm.org}).  A standard fourth-order Shapiro filter is applied to the horizontal momentum, thermodynamics and tracer equations.  For models with fixed rotation period of 2.43 days, we use a horizontal resolution of C32 (equivalent to 128$\times$64 in longitude and latitude). Models with faster rotation rates use  higher horizontal resolution depending on the rotation period. The model domain starts from 100 bars at the bottom to $10^{-3}$ bar at the top, and is divided into 50 vertical levels evenly spaced in log-pressure.

\section{GCM results: comparing simulations with and without hydrogen dissociation and recombination} \label{sec:results}
\subsection{Atmospheric Circulation}
\label{sec:circulation}
\subsubsection{Fixed rotation}
\label{sec:fixrotation}

We first present our GCM results for four sets of simulations with varying equilibrium temperature, $T_{\rm{eq}}$, from 1600 K to 3600 K with in steps of 200 K and a fixed rotation period of 2.43 days. These four sets include models with or without the effects of hydrogen dissociation and recombination included, and with weak drag ($\tau_{\rm{drag}}=10^7$ s) or strong drag ($\tau_{\rm{drag}}=10^4$ s).

\begin{figure*}
\centering
\includegraphics[width=1\textwidth]{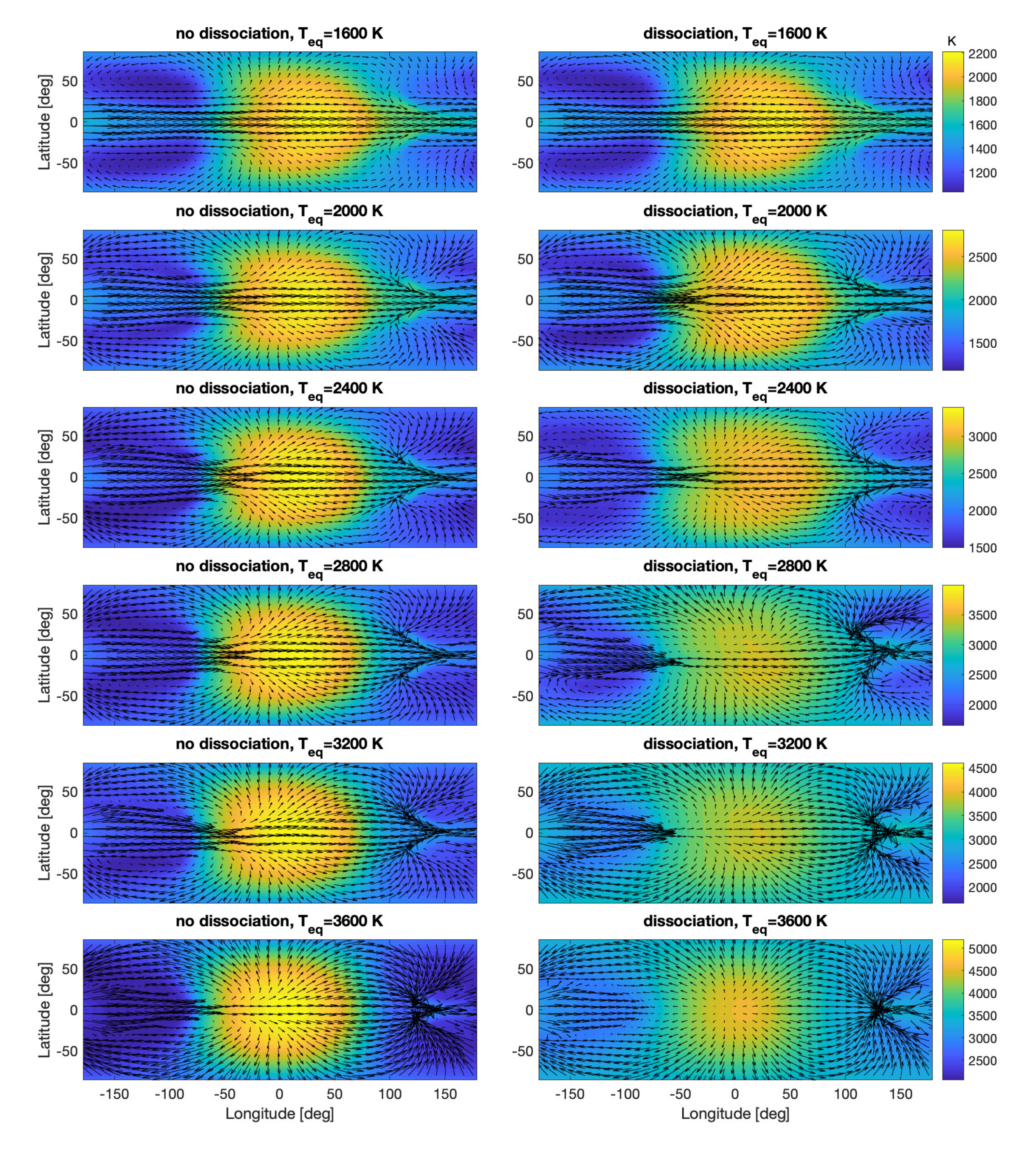}
\caption{Temperature (colors) and winds (vectors) at a pressure of 70 mbar from simulations of hot Jupiter atmospheres with a weak  frictional drag timescale of $10^7~\mathrm{s}$ and varying equilibrium temperature from 1600 K to 3600 K. Rotation period is 2.43 days for all models. The left hand panels show simulations without including the effects of hydrogen dissociation and recombination and the right hand panels show simulations including the effects of hydrogen dissociation and recombination. We find that at relatively high equilibrium temperature,  the hydrogen dissociation cools the dayside of the planet and recombination of atomic hydrogen warms the nightside of the planet, reducing the day-to-night temperature contrast and changing the structure of winds. These effects are more prominent at higher equilibrium temperature. }
  \label{fig:tempmap_d7}
\end{figure*}

\begin{figure*}
\centering
\includegraphics[width=1\textwidth]{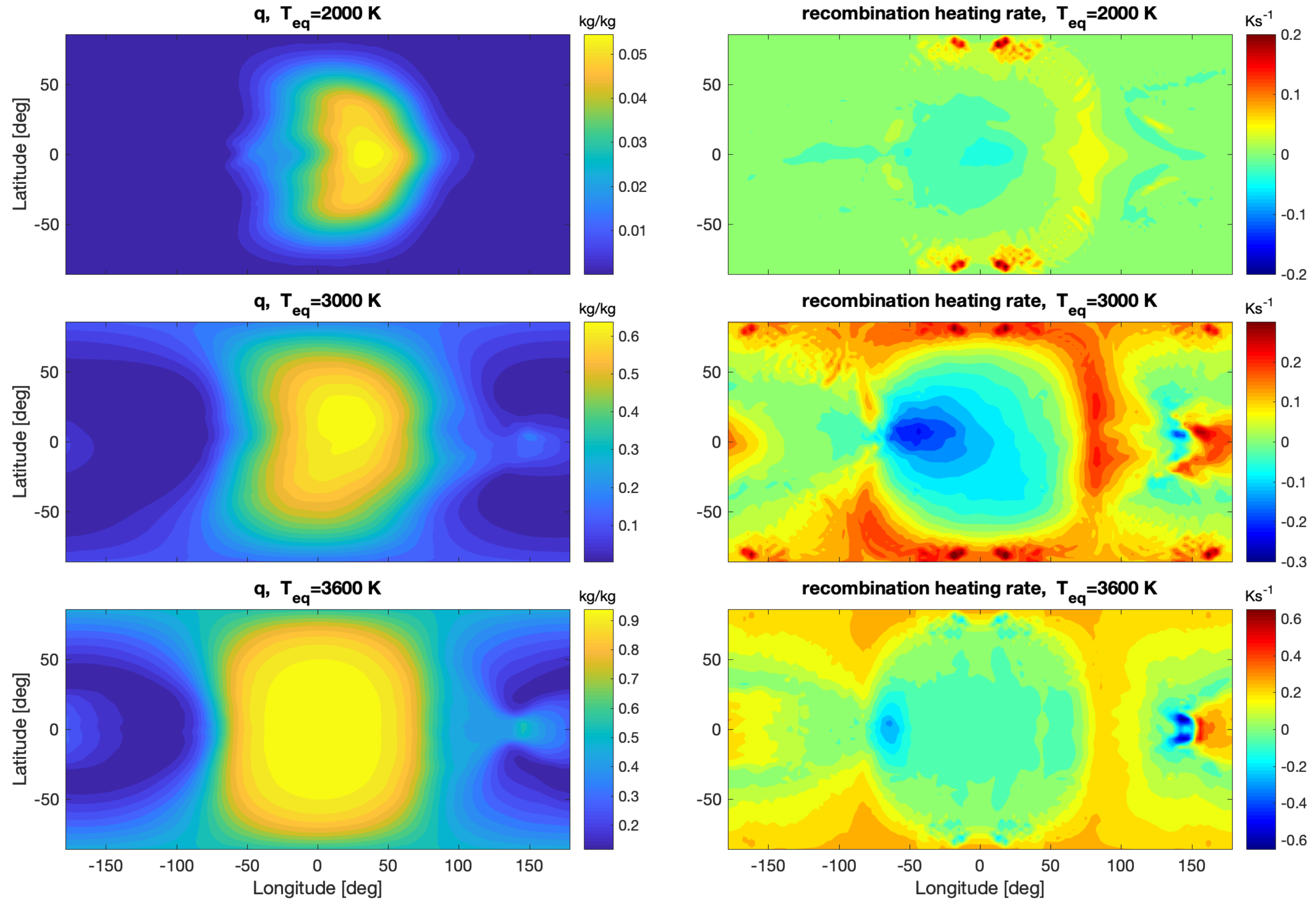}
\caption{Mass mixing ratio of atomic hydrogen (left column) and dissociation and recombination heating rate (right column) from simulations with $T_\mathrm{eq} = 2000 \ \mathrm{K}$ (first row), $T_\mathrm{eq} = 3000 \ \mathrm{K}$ (second row) and $T_\mathrm{eq} = 3600 \ \mathrm{K}$ (third row). These models have weak drag ($\tau_{\rm{drag}}=10^7$ s) and a fixed rotation period 2.43 days. All results are shown at a pressure of $70 \ \mathrm{mbar}$. We find that dissociation occurs largely on the dayside and that the mass mixing ratio of atomic hydrogen is larger in hotter simulations. Dissociation cooling is strong on the dayside for all cases and their spatial patterns correlate well with that of the horizontal gradient of the fractional atomic hydrogen. The recombination heating occurs largely at the limb in the $T_\mathrm{eq} = 2000 \ \mathrm{K}$ simulation, and occurs at both the limb and on the equatorial nightside in the hotter simulations with $T_\mathrm{eq} = 3000 \ \mathrm{K}$ and 3600 K.}
  \label{fig:heating}
\end{figure*}

At relatively low equilibrium temperature ($T_{\rm{eq}} \lesssim 2000$ K), hydrogen dissociation and recombination play a limited role in day-night heat transport. Figure \ref{fig:tempmap_d7} shows the horizontal temperature maps (colors) overlapped with horizontal winds (arrows) at a pressure of about 70 mbar, which is close to the thermal emission level in our GCMs. We show results from simulations with weak drag ($\tau_{\rm{drag}}=10^7$ s) and varying equilibrium temperature from 1600 K to 3600 K. The left column contains the set of models without hydrogen dissociation and recombination included and the right column contains the set of models that include hydrogen dissociation and recombination. At $T_{\rm{eq}} = 1600$ K, the atomic hydrogen fraction is almost negligible everywhere in the atmosphere, and thus the two models with $T_{\rm{eq}} = 1600$ K do not show a recognizable difference in either temperature or wind structures at 70 mbar. At $T_{\rm{eq}} = 2000$ K, the substellar region of the model with hydrogen dissociation is cooler than that without hydrogen dissociation by about 200 K, but the hot region extends further on the dayside for the model with hydrogen dissociation. Near the substellar point, the atomic hydrogen fraction reaches a few percent at 70 mbar (shown in Figure \ref{fig:heating}). Winds transport the fractional atomic hydrogen out of the substellar region to  cooler surroundings, and the recombination of hydrogen warms up these areas. By mass conservation, molecular hydrogen is transported toward the substellar point and part of it turns into atomic hydrogen and cools the substellar area. However, because the region over which hydrogen dissociation is strong is small, the heating/cooling effect on heat redistribution is mostly confined on the dayside. As a result, at $T_{\rm{eq}} = 2000$ K the heating/cooling effect of hydrogen recombination and dissociation has a negligible contribution to the day-night heat transport.   

At $T_{\rm{eq}} = 2400$ K, dissociation of molecular hydrogen on the dayside starts to substantially cool the dayside atmosphere and the recombination of atomic hydrogen warms the terminator regions. Although the recombination of hydrogen does not reach to the nightside, the extra heat carried by hydrogen recombination to the terminator is significant. The distance over which circulation is required to transport heat from day to night is therefore shortened. As a result, the day-night heat transport in our simulations with $T_{\rm{eq}} = 2400$ K that include hydrogen dissociation and recombination is more efficient than simulations does not include this effect. 

The amount by which hydrogen dissociation and recombination reduces the day-night temperature contrast increases with increasing equilibrium temperature, and this can be readily recognized by reading the color scale of the temperature map in Figure \ref{fig:tempmap_d7}. We will quantify the day-night heat transport further in Section \ref{sec:phasecurve}. The dayside atomic hydrogen fraction as well as the day-night fractional atomic hydrogen difference increases with increasing equilibrium temperature, and thus the cooling of the dayside and increased day-night heat transport due to hydrogen dissociation becomes more prominent as the equilibrium temperature increases.  This trend starts from an equilibrium temperature of $\sim$2200 K and continues up to $\sim$3200 K. 

\begin{figure}
    \centering
    \includegraphics[width=0.5\textwidth]{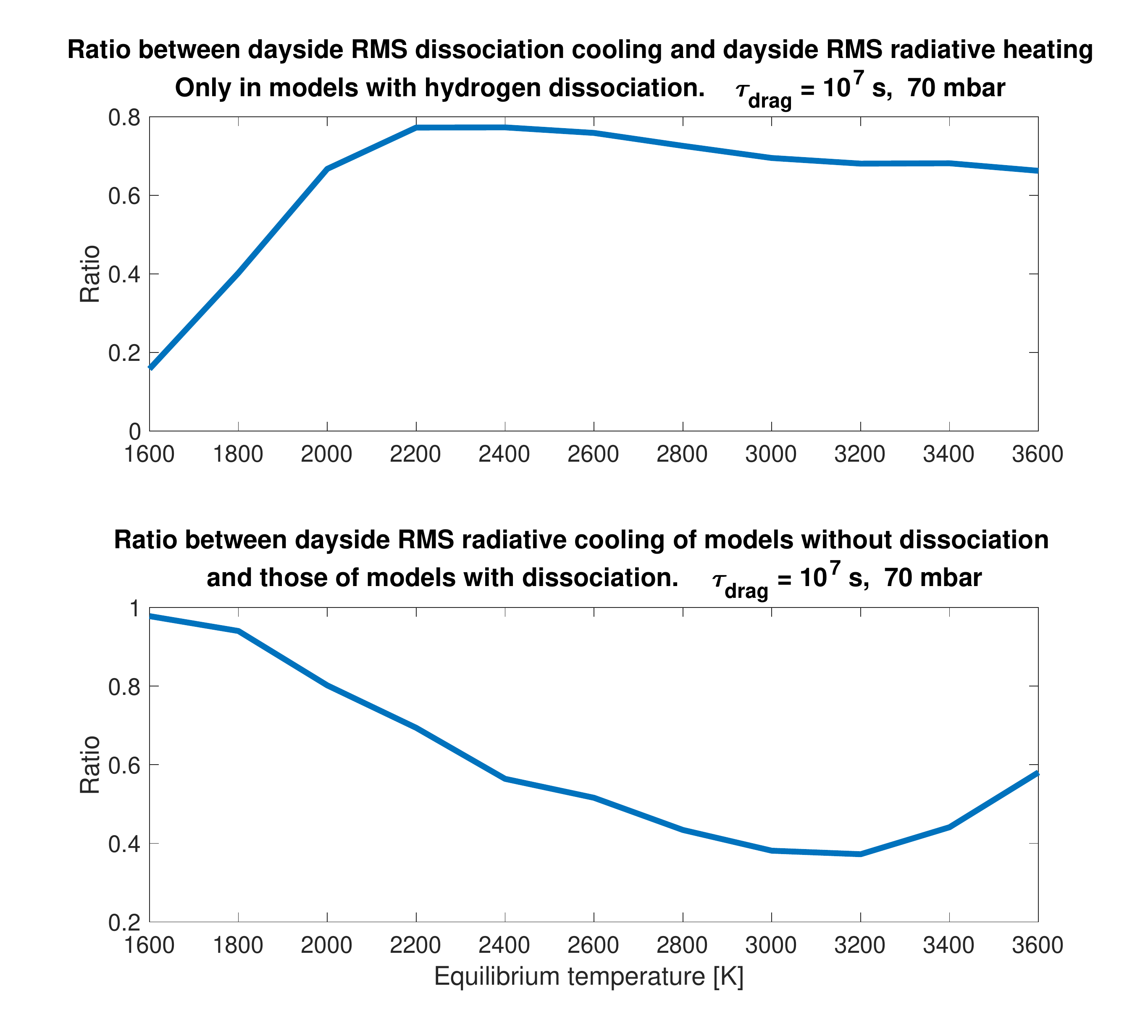}
    \caption{Top: The ratio between the RMS of the dayside radiative heating rate and the RMS of dayside hydrogen dissociation cooling rate at 70 mbar for the weak-drag simulations including hydrogen dissociation shown in Figure \ref{fig:tempmap_d7}. Bottom: The ratio between the RMS of the dayside radiative heating rate in simulations without hydrogen dissociation and those with hydrogen dissociation. In both panels, the RMS heating/cooling rates are taken between $\pm80^{\circ}$ in longitude and $\pm40^{\circ}$ in latitude on the dayside.}
    \label{fig:heatingrate}
\end{figure}

However, the trend of decreasing fractional day-night temperature contrast with increasing equilibrium temperature  weakens or even reverses for equilibrium temperatures higher than 3200 K. At sufficiently high equilibrium temperature the hydrogen dissociation fraction is saturated on the dayside, and both the horizontal or vertical gradient of fractional atomic hydrogen over the dayside become small. As a result, the cooling effect of dissociation on the dayside is no longer as efficient as that in slightly cooler atmospheres. Instead, radiative damping starts to drive the fractional day-night temperature contrast to a higher value for planets with equilibrium temperatures higher than 3200 K.

The heating/cooling budget by hydrogen recombination/dissociation becomes increasingly important in the thermodynamic balance of hot atmospheres \citep{Komacek:2018aa}. This is characterized by the ratio between the root-mean-square (RMS) of the dayside radiative heating rate and the RMS of dayside hydrogen dissociation cooling rate. We show in the upper panel of Figure \ref{fig:heatingrate} this ratio at 70 mbar for the weak-drag simulations including hydrogen dissociation shown in Figure \ref{fig:tempmap_d7}. The result is averaged over the dayside area between $\pm 80^{\circ}$ in longitude and $\pm 40^{\circ}$ in latitude. Somewhat surprisingly, this ratio is not small even at $T_{\rm{eq}}=1600$ K and quickly increases with increasing equilibrium temperature, then saturates at a value between 65$\%$ and 80$\%$ starting from $T_{\rm{eq}}=2200$ K, where the day-night temperature differences just begin affected noticeably by hydrogen dissociation. This ratio is always less than one because the dayside heating by radiation are balanced by heat transport of both the heat budget of fractional atomic hydrogen $\mathcal{L}_h q$ and the dry air enthalpy $\bar{c_p}T$. \\
\indent Neglecting the frictional dissipation of heat and the change of $\bar{c_p}$, one may rewrite the thermodynamic Equation (\ref{eq.thermo}) as 
\begin{equation}
    \frac{d}{dt}(\bar{c_p}T+\mathcal{L}_h q) - \frac{\omega}{\bar{\rho}} = g\frac{\partial F}{\partial p}.
    \label{eq.thermo2}
\end{equation}
The advection of the so-called ``moist enthalpy'', $\bar{c_p}T+\mathcal{L}_h q$, together with decompressional cooling, balances the radiative heating. As shown in the upper panel of Figure \ref{fig:heatingrate}, the cooling budget on the dayside is dominated by the transport of $\mathcal{L}_h q$ at $T_{\rm{eq}}\gtrsim 2200$ K. 
The ratio between the dayside RMS radiative heating rate in simulations without hydrogen dissociation and that in simulations including hydrogen dissociation at 70 mbar is shown in the lower panel of Figure \ref{fig:heatingrate}. This ratio quantifies the dayside cooling effect by including hydrogen dissociation.  Consistent with the upper panel in Figure \ref{fig:heatingrate}, this ratio is close to but less than one at low equilibrium temperature, and gradually decreases with increasing equilibrium temperature until it reaches a value of $\approx 0.4$ at $T_{\rm{eq}}\sim 3200$ K, after which it increases with increasing equilibrium temperature. As discussed above, in the hottest atmospheres we consider ($T_{\rm{eq}}\gtrsim 3200$ K), the dayside hydrogen dissociation is saturated, and advection of dry enthalpy becomes increasingly important to balance radiative heating. As a result, the hottest hot Jupiters exhibit a similar trend of increasing day-night temperature contrast with increasing equilibrium temperature as found for cooler hot Jupiters \citep{Perez-Becker:2013fv,Komacek:2015}.

To better illustrate the horizontal distribution of fractional atomic hydrogen and its heating/cooling effects on the atmosphere, we show in the left hand column of Figure \ref{fig:heating} the  horizontal maps of the mass mixing ratio of atomic hydrogen at 70 mbar. The upper panel shows results from the weak-drag simulation with $T_{\rm{eq}}=2000$ K, the middle panel shows results for the simulation with $T_{\rm{eq}}=3000$ K and the lower panel shows results for the simulation with $T_{\rm{eq}}=3600$ K. The corresponding cooling and heating rates caused by hydrogen dissociation and recombination are shown in the right hand column of Figure \ref{fig:heating}. The spatial distribution of the atomic hydrogen fraction is very close to chemical equilibrium as a result of the short relaxation timescale in our models. The simulation with a cool equilibrium temperature of 2000 K contains only a few percent of atomic hydrogen on the dayside, with almost no atomic hydrogen on the nightside. As a result, the transport of atomic hydrogen in the cool simulation is confined within the dayside and most heating caused by hydrogen recombination occurs near the terminator, similar to the results of \cite{Bell:2018aa}.  
In the hot cases with $T_{\rm{eq}}=3000$ and 3600 K, a high fraction of atomic hydrogen extends from dayside to nightside, through both high latitudes and the eastern equatorial region. The heating rate due to hydrogen recombination can then extend to the nightside and provides direct heating. Interestingly, although the atomic hydrogen fraction at $T_{\rm{eq}}=2000$ K is less than a tenth of that at $T_{\rm{eq}}=3000$ and 3600 K, the recombination heating rates of the all three cases are on the same order. This is probably due to the compensation by weaker winds in the hot simulations. Indeed, the RMS of horizontal winds near 70 mbar for weak-drag simulations including hydrogen dissociation is noticeably smaller than that for simulations without hydrogen dissociation, by up to 1000 $\rm{ms^{-1}}$ in the hot simulations (not shown).

\begin{figure*}
\centering
\includegraphics[width=1\textwidth]{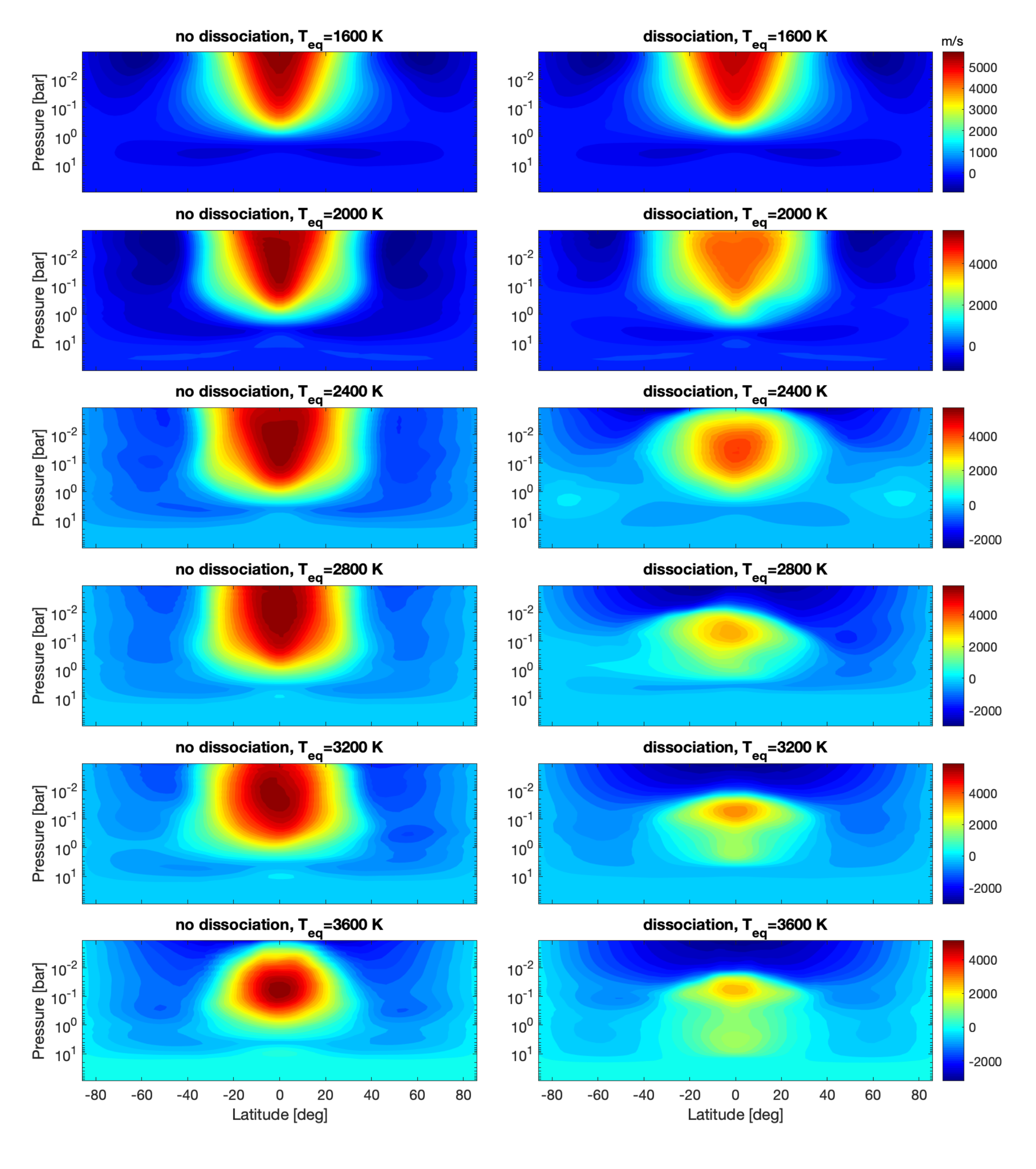}
\caption{Zonal-mean zonal wind speeds as a function of latitude and pressure from GCM simulations with a weak fixed frictional drag timescale of $10^7~\mathrm{s}$, a fixed rotation period 2.43 days, and varying equilibrium temperature from $1600~\mathrm{K}$ to $3600~\mathrm{K}$. The left-hand panels show zonal-mean zonal winds without including the effects of hydrogen dissociation and recombination, and the right hand panels include this effect in the GCM. We find that including hydrogen dissociation and recombination reduces the strength of the equatorial zonal jet of ultra-hot Jupiters.}
  \label{fig:uzonal_d7}
\end{figure*}

 We are also interested in the effects of hydrogen dissociation and recombination on the detailed circulation pattern. At a first glance, the horizontal wind fields at 70 mbar shown in Figure \ref{fig:tempmap_d7} are qualitatively similar between simulations without and with hydrogen dissociation, and most of them exhibit an eastward equatorial jet with the characteristic off-equatorial Rossby waves and equatorial Kelvin waves driven by day-night forcing (e.g., \citealp{Showman_Polvani_2011,Tsai:2014,Hammond:2018aa}). However, at very high equilibrium temperature ($T_{\rm{eq}} \gtrsim 3200$ K), the eastward equatorial jets at 70 mbar in simulations with hydrogen dissociation are significantly weakened and even disappears at $T_{\rm{eq}}=3600$ K, while those in simulations without hydrogen dissociation are still well preserved. 
 
 More significant differences in the wind field can be better recognized in  the time-averaged, zonal-mean zonal wind profiles  shown in Figure \ref{fig:uzonal_d7} for models with weak drag and a fixed rotation period.  Two main features are interesting. First, the overall speeds of the eastward equatorial jets are weaker in the models with hydrogen dissociation than those of models without hydrogen dissociation. This feature starts even at  an  equilibrium temperature of 1600 K, in which the shape of the jets are almost identical but the  magnitude of the jet in the dissociation model is slightly smaller. Starting from equilibrium temperature of 2000 K, the reduction of the jet speed in simulations with hydrogen dissociation is prominent. The speeds of the eastward jet cores in high-temperature simulations including hydrogen dissociation and recombination are only a fraction of those of simulations without hydrogen dissociation and recombination. This may be partly understood as the reduction of horizontal eddy winds in models including hydrogen dissociation, which are essential to pump eastward angular momentum to the equatorial region. As shown by Equation (\ref{eq.thermo2}) and \cite{Komacek:2018aa}, the enthalpy budget $\bar{c_p}T+\mathcal{L}_h q$ can be significantly enhanced with a modest amount of atomic hydrogen. As a result, weaker winds are needed to balance the overall day-night thermal forcing at high equilibrium temperature. 

The second, even more puzzling feature of the zonal-mean zonal winds is that at lower pressures, models including hydrogen dissociation exhibit equatorial westward flows at equilibrium temperature higher than 2400 K. The magnitude and the vertical extent of the westward zonal-mean equatorial winds increase with increasing equilibrium temperature. In models without hydrogen dissociation, the eastward equatorial jets are strong and robust all the way up to the upper model boundary for all equilibrium temperatures. This is puzzling because the eastward equatorial jet is likely a robust outcome of many hot-Jupiter GCMs (see a review in \citealp{Heng:2014b}). To understand the role of eddy winds, we performed an analysis of the zonal-mean eddy angular momentum transport, and found that in models including hydrogen dissociation, the horizontal eddy transport at low pressures exert an eastward forcing on the equatorial jet. This eastward forcing is largely balanced by a westward forcing caused by vertical eddy transport in a statistically equilibrium state. The diagnosis in models without hydrogen dissociation shows a similar picture. This suggests that the horizontal eddies do not directly cause the westward equatorial flow.  

 The balance between horizontal and vertical eddy transport at low latitudes is likely a necessary consequence in statistical equilibrium, and thus the above diagnosis does not provide insight on why the eastward equatorial jet is (partially) suppressed. To gain further insight, we performed the following two analyses. First, we diagnosed the spin-up phase of simulations with $T_{\rm{eq}} = 3200$ K both with and without hydrogen dissociation and recombination. In the spin-up phase, the vertical eddy transport in both simulations exerts a westward forcing on the equatorial flow and the horizontal eddy transport exerts an eastward forcing. However, in the case with hydrogen dissociation, the overall westward forcing caused by vertical eddy transport is slightly stronger than the eastward forcing by horizontal eddy transport at low pressures, and the model quickly develops a westward equatorial flow at low pressure. On the other hand, in the case without hydrogen dissociation the forcing due to horizontal eddy transport overcomes that of vertical eddy transport and drives a eastward equatorial flow. Second, we took the equilibrium atmospheric state of a model without hydrogen dissociation, which has a strong eastward equatorial jet and a balanced eddy angular momentum forcing, and gradually turn on the effects of hydrogen dissociation. Similar to the exercise of the spin-up phase, the westward equatorial forcing by vertical eddy transport slightly exceeds the eastward forcing by horizontal eddy transport once the hydrogen dissociation is turned on, and eventually the model transitions to a state with westward equatorial zonal flow at low pressures. 

The above two exercises demonstrate the robustness of the westward flow at low pressure in hot models including hydrogen dissociation and with a fixed rotation period (2.43 days), and also demonstrate the role of vertical eddy transport. However, at this point, the mechanism of vertical-eddy-driven westward equatorial flow, as well as why this occurs at high equilibrium temperature when hydrogen dissociation is included, are still unclear.
Ultra-hot Jupiters with sufficiently high equilibrium temperature likely have shorter rotation periods than that (2.43 days) of models shown in Figure \ref{fig:uzonal_d7}, and, as will be shown below, the westward equatorial flow is not a persistent outcome in faster rotating models. Although interesting, understanding the origin of westward equatorial flow at low pressure is not urgent, and a more careful exploration of why and under what conditions the westward equatorial flow would occur is deferred for future work. 

\begin{figure*}
\centering
\includegraphics[width=1\textwidth]{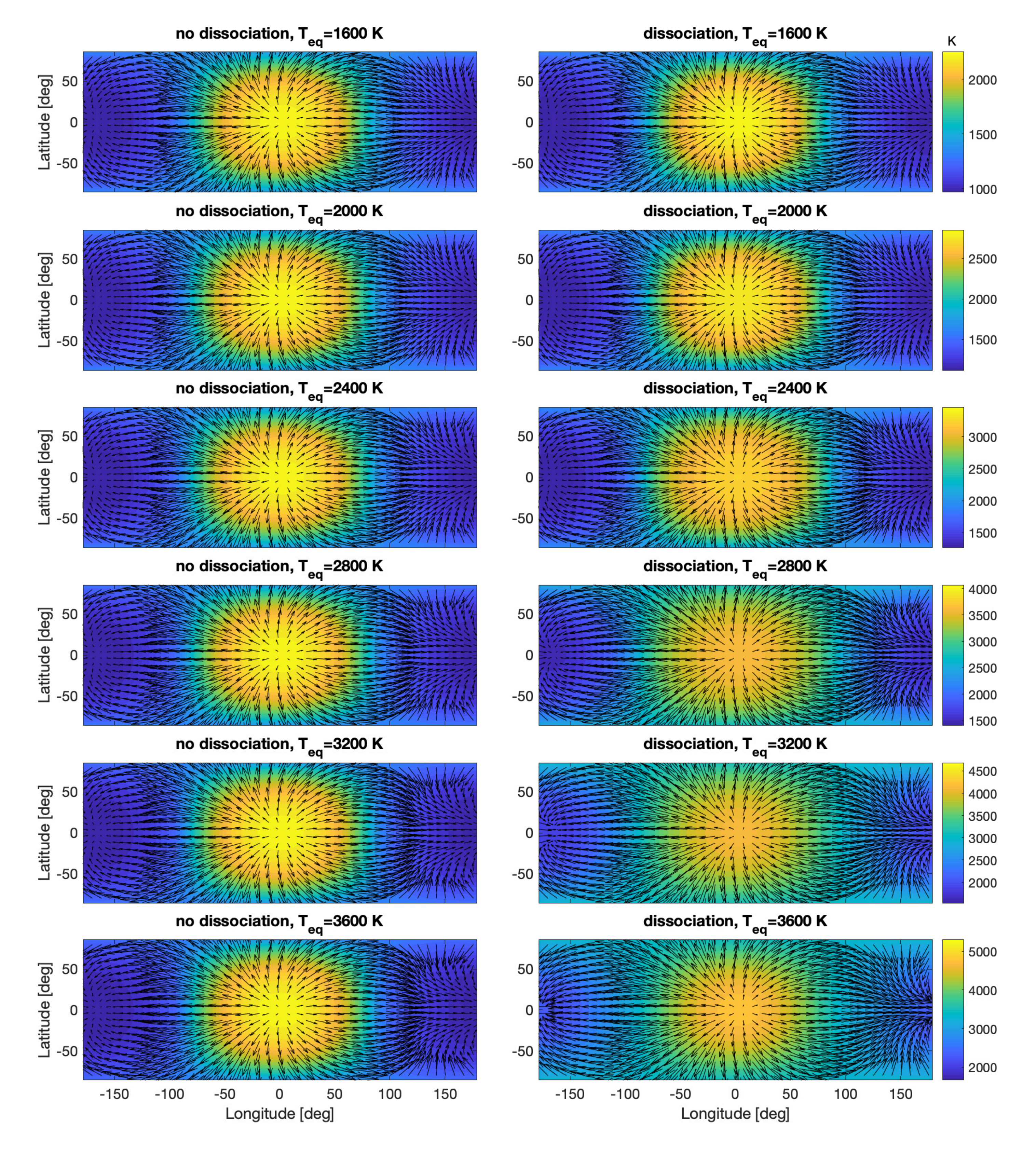}
\caption{Temperature (colors) and winds (vectors) at a pressure of 70 mbar from simulations of hot Jupiter atmospheres with a strong fixed frictional drag timescale of $10^4~\mathrm{s}$, a fixed rotation period 2.43 days, and varying equilibrium temperature from 1600 K to 3600 K. The left hand panels show simulations without including the effects of hydrogen dissociation and recombination and the right hand panels show simulations including the effects of hydrogen dissociation and recombination. The circulation of all models are dominated by a day-to-night flow. As in \Fig{fig:tempmap_d7}, we find that including hydrogen dissociation and recombination into our simulations reduces the day-to-night temperature contrast of ultra-hot Jupiters.}
  \label{fig:tempmap_d4}
\end{figure*}

Magnetohydrodynamic drag could be significant in the atmospheres of ultra-hot Jupiters due to substantial thermal ionization. This effect is crudely represented using strong frictional drag in our models. Figure \ref{fig:tempmap_d4} shows horizontal temperature maps at 70 mbar overlapped with horizontal wind vectors for two sets of models with different equilibrium temperature but with strong drag ($\tau_{\rm{drag}}=10^4$ s). The atmospheric circulation of these models  are all dominated by a day-to-night flow that is roughly symmetric about the substellar and anti-substellar points. This is because the drag timescale is shorter than the rotational timescale, and the balance in the horizontal angular momentum is primarily between the day-night thermal forcing and the drag force, giving rise to day-night flow \citep{Showman_Polvani_2011,Komacek:2015,Komacek:2017}. Wave motions are effectively damped as the wave propagation timescale is longer than the drag timescale.  Even in the strong-drag regime, the heating and cooling effect by hydrogen recombination and dissociation plays a crucial role in reducing the day-night temperature contrast for equilibrium temperatures hotter than $\sim 2400$ K. This is not surprising, as the heat budget carried by hydrogen dissociation and recombination becomes large at high atmospheric temperatures where the fraction of atomic hydrogen becomes large. Even a small amount of fractional atomic hydrogen transport can help to further reduce the day-night temperature contrast in addition to the advection of dry air enthalpy.


\subsubsection{Varying rotation}
We investigate the effect of varying rotation rate on the general circulation of ultra-hot Jupiters in tandem with varying equilibrium temperature. This is done by adjusting the rotation period of the planet at the value equal to its orbital period given an arbitrary equilibrium temperature of the planet and a set of fixed stellar parameters as follows (e.g., \citealp{Komacek:2017}):
\begin{equation}
P_{\rm{rot}} =  \frac{\pi}{\sqrt{2}} \left( \frac{T_{\star}}{T_{\rm{eq}}} \right)^3  \sqrt{\frac{R_{\star}^3}{GM_{\star}} } ,
\end{equation}
where $P_{\rm{rot}}$ is the planetary rotation period, $T_{\star}$ is the stellar effective temperature,  $R_{\star}$ is the stellar radius, $M_{{\star}}$ is the stellar mass and $G$ is the gravitational constant. As mentioned in Section \ref{sec.numerical}, we assume a star with $T_{{\star}}=6000$ K, $R_{{\star}}=1.8$ solar radii, and $M_{{\star}}$ = 1.2 solar mass. With $T_{\rm{eq}}=2000$ K, the planetary orbital (set equal to the rotational) period is 2.43 days, the same as that of the main suite of simulations discussed in Section \ref{sec:fixrotation}. In the set of models with effective temperature $T_{\rm{eq}}=2200$, 2400, 2600, 2800, 3000 and 3200 K, the corresponding rotation periods are  1.83, 1.41, 1.11, 0.89, 0.72, and 0.59 days, respectively.

\begin{figure*}
\centering
\includegraphics[width=1\textwidth]{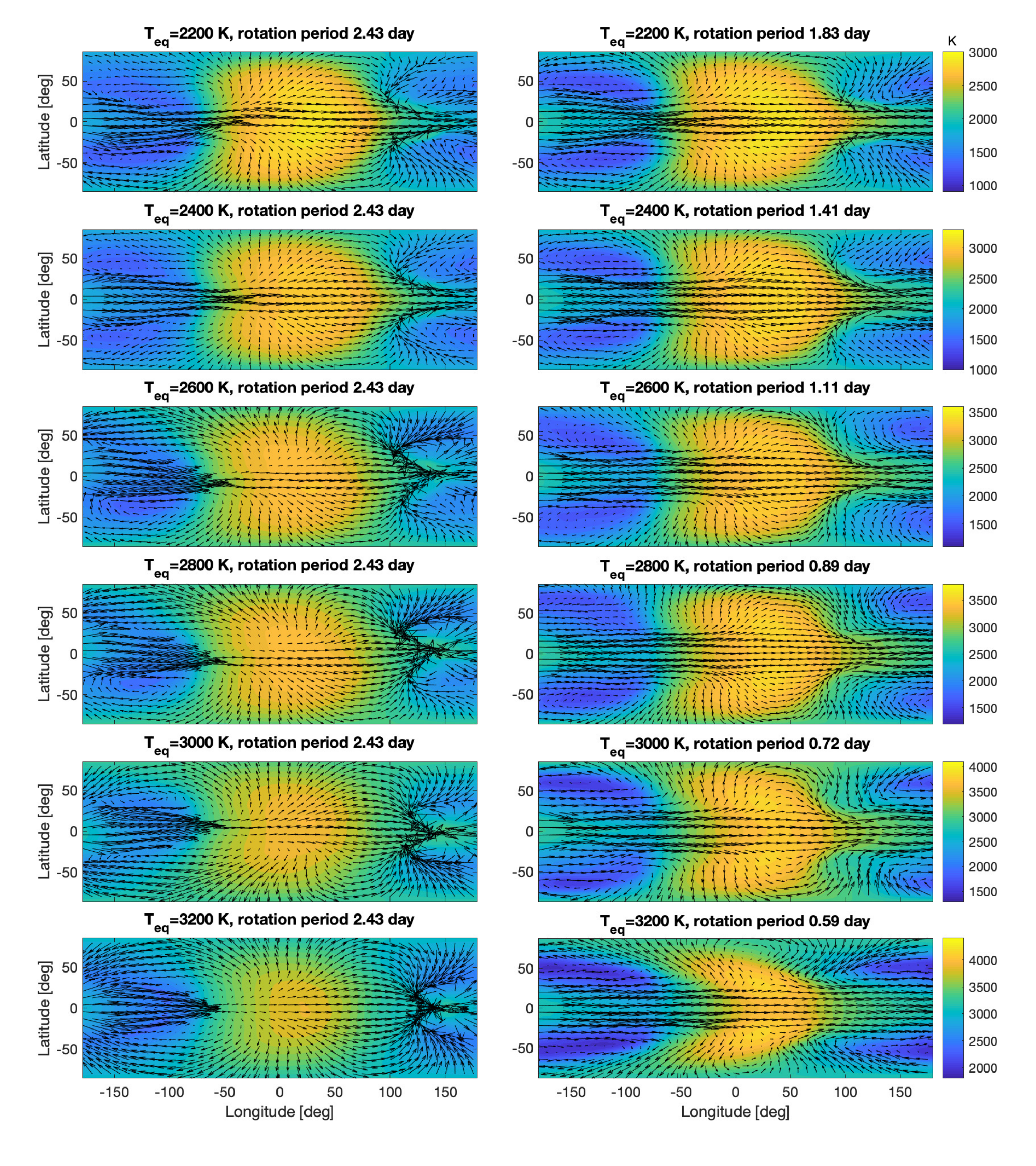}
\caption{Temperature (colors) and wind (arrows) maps from GCM simulations with a fixed rotation period (left column) and consistently decreasing rotation period with increasing equilibrium temperature (right column). All simulations include weak drag and the effects of hydrogen dissociation and recombination. Due to the fast rotation at high equilibrium temperatures, the dayside is warmer and the nightside is cooler in simulations including a varying rotation period.}
  \label{fig:temp_d7_rotconsis}
\end{figure*}

We compare results from simulations with the same equilibrium temperatures, weak drag ($\tau_{\rm{drag}}=10^7$ s), and all including hydrogen dissociation but with different rotation periods, in Figure \ref{fig:temp_d7_rotconsis}. Figure \ref{fig:temp_d7_rotconsis} shows horizontal temperature maps at 70 mbar for the set of simulations with varying equilibrium temperature but with a fixed rotation period of 2.43 days on the left column, and simulations with varying equilibrium temperature and varying rotation period on the right column. The rotation period of each simulation is shown above each panel. 
In general, winds at high latitudes of faster rotators with rotation periods less than a day tend to be slightly closer to the geostrophic balance (the balance between pressure gradient force and the Coriolis force), such that the velocity vectors are more parallel to isotherms at isobaric surfaces. As a result, high-latitude temperature anomalies in rapidly rotating simulations are slightly larger than those with a fixed rotation period due to the stronger effect of rotation. Most obviously, the nightside Rossby gyres of rapid rotators have lower temperatures than those with a fixed rotation period. Overall, the rapidly rotating models exhibit more robust and stronger eastward equatorial jets than those with a fixed rotation period, in the sense that winds at the equator are more zonally uniform. Because of the stronger jets and thus a shorter advective timescale, the rapid rotators have smaller  day-night temperature contrasts at low latitudes. This compensates for the larger horizontal temperature anomalies at high latitudes. The competition between the equatorial and high-latitude processes determines the overall day-night temperature difference, which will be discussed in Section \ref{sec:phasecurve}. The strong eastward equatorial jets in rapid rotators also result in greater eastward shifts of the wave patterns, and thus could lead to a larger phase offset in their phase curves. 

  We expect the meridional extent of the wave patterns to decrease with decreasing rotation period due to the smaller equatorial deformation radius \citep{Showman_Polvani_2011}. This has been numerically confirmed by Tan \& Showman (2019, in prep) in the context of Newtonian cooling models. However, somewhat surprisingly, our simulations here show no obvious decrease of the meridional extent of the waves with decreasing rotation period, except in the case with the shortest rotation period of 0.59 days in which the thermal pattern is noticeably smaller in latitudinal extent compared to the case with a slightly longer rotation period of 0.72 days. It is likely that the wave dynamics depends on various other parameters in addition to rotation, which could all be affected by changing the equilibrium temperature.   Indeed, in the context of shallow water systems \citep{Showman_Polvani_2011,Hammond:2018aa}, the forced-damped wave solutions are affected  by the radiative forcing, frictional drag, and rotation, all of which determine the effective deformation radius. Including effects of hydrogen dissociation could only increase the complexity of wave dynamics. Further theoretical investigation of equatorial waves including hydrogen dissociation is required to gain a fundamental understanding of waves and the wave-mean-flow interactions in atmospheres of ultra-hot Jupiters.

\begin{figure*}
\centering
\includegraphics[width=1\textwidth]{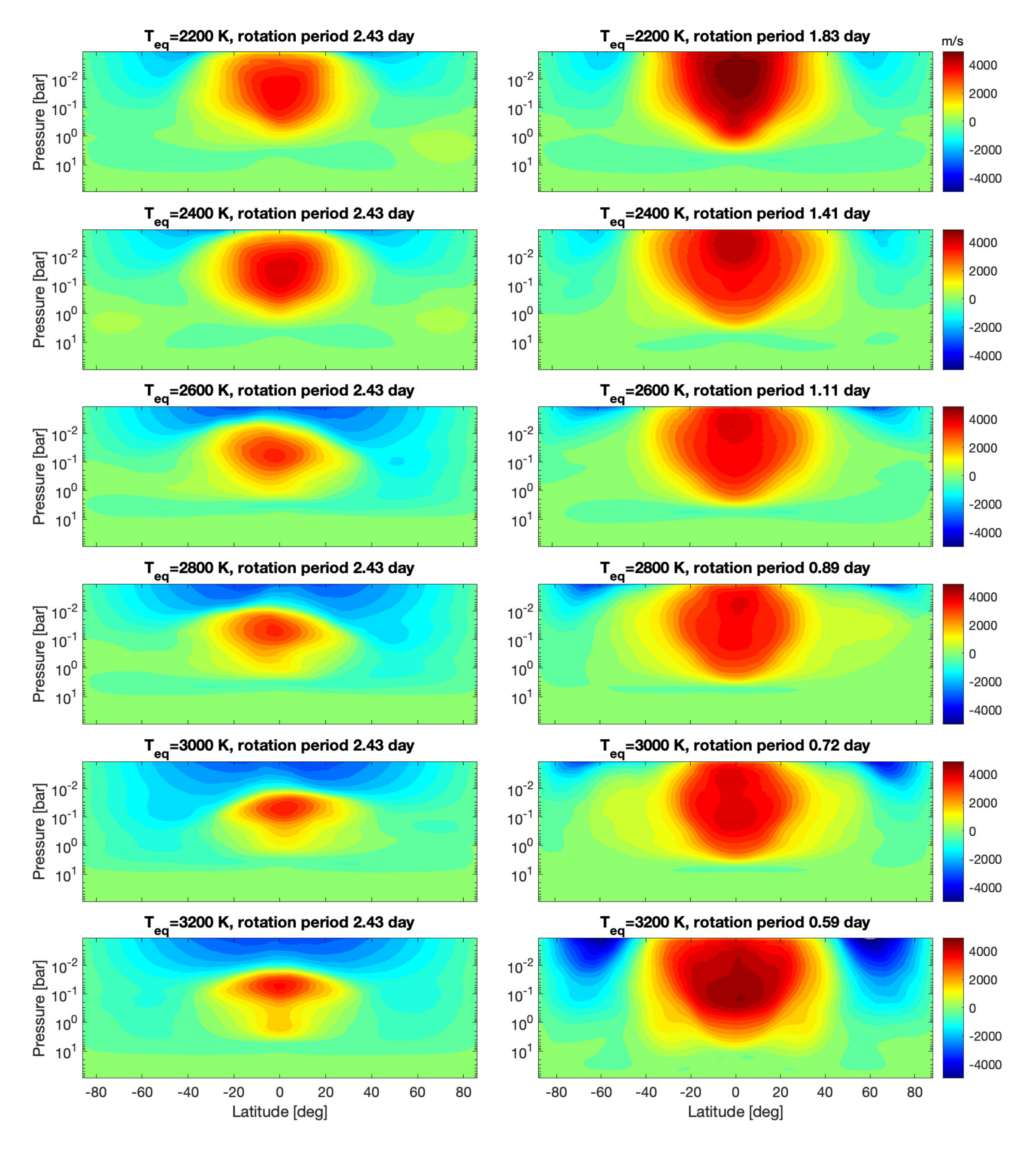}
\caption{Zonal-mean zonal wind speeds from GCM simulations with a fixed rotation period (left column) and consistently decreasing rotation period with increasing equilibrium temperature (right column). All simulations include weak drag and the effects of hydrogen dissociation and recombination. We find that there is an equatorial superrotating jet throughout the observable atmosphere when including the decreasing rotation period with increasing equilibrium temperature.}
  \label{fig:uzonal_d7_rotconsis}
\end{figure*}

The zonal-mean zonal wind profiles of simulations with varying rotation period are in sharp contrast with those that have a fixed rotation period, as shown in Figure \ref{fig:uzonal_d7_rotconsis} for the same sets of models in Figure \ref{fig:temp_d7_rotconsis}. The rapid rotators all exhibit strong and robust eastward equatorial jets, and the eastward equatorial jets extend from several bars up to the upper model boundary. Meanwhile, the equatorial jets in simulations with a fixed rotation period weaken with increasing equilibrium temperature, and exhibit westward equatorial flow at low pressure. The meridional extent of the eastward equatorial jet in faster rotators is not sensitive to rotation period. This is probably because, as shown in Figure \ref{fig:temp_d7_rotconsis}, the meridional extent of horizontal wave patterns does not change with decreasing rotation period, and so the resulting wave-mean-flow interactions naturally leads to the invariant meridional jet width. A possible explanation of the difference in the zonal-mean zonal winds between the two sets  of  simulations  is  that  the  increasing  rotation  rate allows a more robust formation of horizontal wave patterns which efficiently pump eastward angular momentum to the  equatorial flow, and this effect overcomes the effects responsible for the westward equatorial flow at low pressures seen in the fixed-rotation models. 

Our fast rotating models, especially those with $T_{\rm{eq}}\gtrsim 3000$ K and $P_{\rm{rot}}\lesssim 0.72$ day, display variability in both temperature and wind fields. The variability is caused by the appearance and disappearance of high-latitude vortices, with typical size of several tens of degrees in latitude and longitude and typical life time of more than 10 days. The vortices are in the form of either cyclones\footnote{The relative vorticity of a cyclone has the same sign of the local Coriolis parameter, whereas that of an anticyclone has the opposite sign of the local Coriolis parameter.} which typically have a cold core and anticyclones which typically have a warm core. Once they form, the vortices can drift along  latitude bands, causing perturbations on either the dayside or nightside temperature. The origin of the vortices is probably due to the large-scale atmospheric instabilities (either barotropic or baroclinic instability), which are typical in rapidly rotating planetary atmospheres \citep{vallis2017}. At depths where the radiative damping timescale is much longer than relevant dynamical timescales, the basic atmospheric state is characterized by a zonally symmetric configuration \citep{Showman:2014}. The day-night forcing pattern together with the fast rotation enforce a large equator-to-pole temperature difference with associated fast zonal jets, and such a structure could be favored by large-scale instabilities which would lead to the development of large-scale vortices. In our rapid rotating models, the vertical wave length of the vortices is long. Although the origin of vortices is likely at relatively high pressures, the temperature perturbation can penetrate to the photosphere, thereby affecting observable properties of the planet.  We will present the observational consequences of this atmospheric variability in Section \ref{sec:variability}.

\subsection{Phase Curves}
\label{sec:phasecurve}
In this section, we present phase curves from the sets of simulations described above, calculated by post-processing them with our double-grey radiative transfer scheme.

\subsubsection{Fixed rotation}
\begin{figure*}
\centering
\includegraphics[width=1\textwidth]{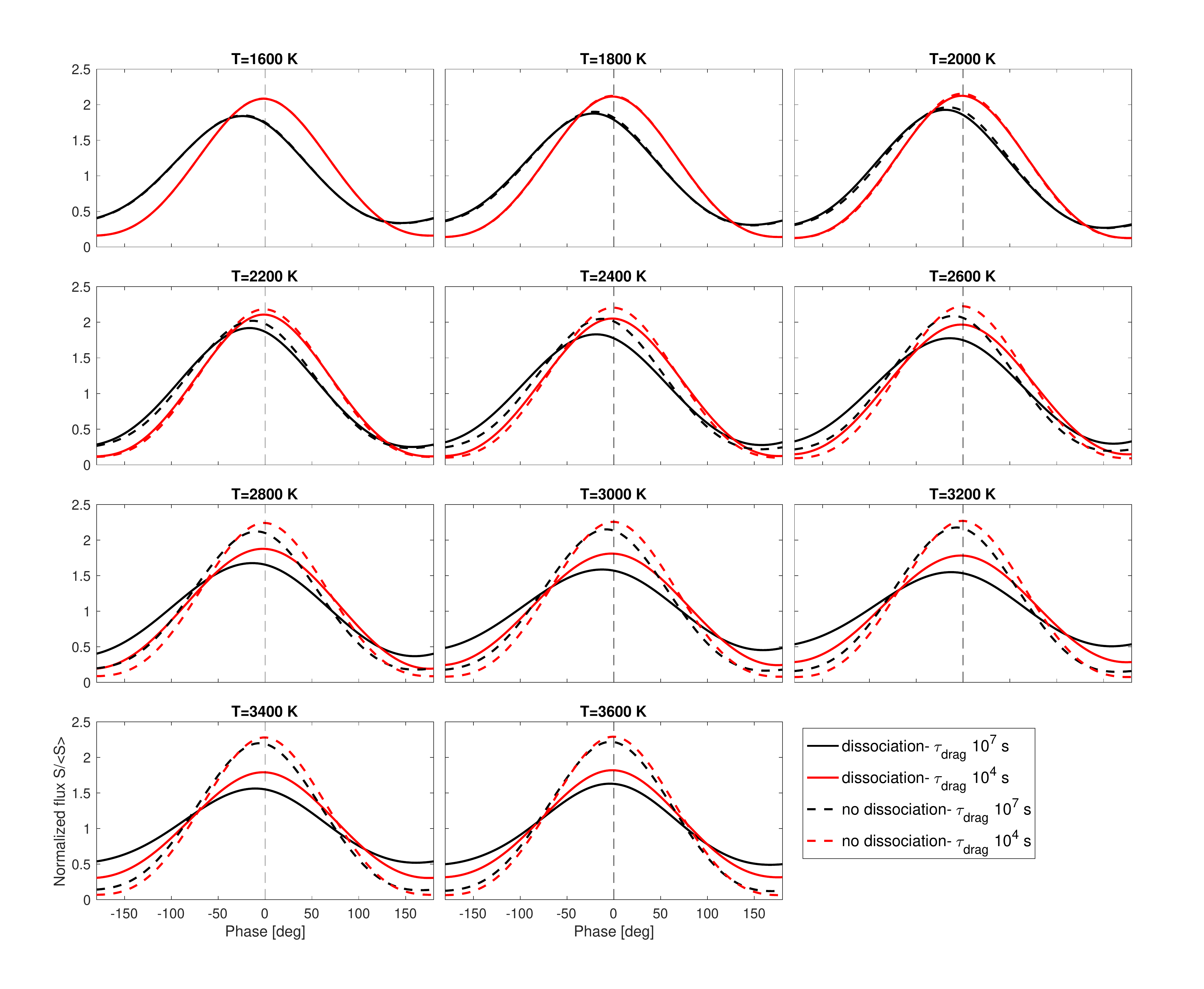}
\caption{Normalized full-phase light curves calculated from our GCM simulations with a fixed rotation period 2.43 days and varying equilibrium temperature from $1600-3600~\mathrm{K}$. Black lines show simulations with weak drag ($\tau_\mathrm{drag} = 10^7 \ \mathrm{s}$), while red lines show simulations with strong drag ($\tau_\mathrm{drag} = 10^4 \ \mathrm{s}$). We normalize each phase curve by the average emitted flux in order to display them on the same scale. The solid lines include the effects of hydrogen dissociation and recombination, while the dashed lines ignore this effect.  We find that including hydrogen dissociation and recombination significantly reduces phase curve amplitudes and increases the phase curve offset of ultra-hot Jupiters, with these effects increasing with increasing equilibrium temperature.}
  \label{fig:phasecurves_d7}
\end{figure*}

\Fig{fig:phasecurves_d7} shows normalized phase curves calculated from our full suite of GCM experiments with a fixed rotation period of 2.43 days. We find that simulations with stronger drag have smaller phase curve offsets and larger phase curve amplitudes, as found in \cite{Komacek:2017}. The differences between phase curves from simulations with and without hydrogen dissociation and recombination increase with increasing equilibrium temperature. Our simulations with $T_\mathrm{eq} = 1600 \ \mathrm{K}$ show almost no difference in the phase curves with and without hydrogen dissociation and recombination. However, our simulations with high $T_\mathrm{eq} = 3600 \ \mathrm{K}$ have a $\sim 50\%$ smaller flux at secondary eclipse when including hydrogen dissociation and recombination, with an even larger relative increase in the nightside flux.  \\
\indent With increasing equilibrium temperature, we find that simulations with hydrogen dissociation and recombination have lower phase curve amplitudes than simulations that do not include their effects. For strong-drag models, no obvious phase offset can be easily recognized in their phase curves. For weak-drag models, those with hydrogen dissociation and recombination have systematically larger phase curve offsets than models without their effects. The reduced phase curve amplitude in our simulations with hydrogen dissociation and recombination is due to the cooler daysides due to dissociation cooling and warmer nightsides from recombination heating that we found in \Sec{sec:circulation}. The enhanced phase curve offsets in our weak-drag simulations with hydrogen dissociation and recombination is interesting. One might expect the other way around because of two reasons. First, the eastward equatorial jets are weaker in models including hydrogen dissociation (see Figure \ref{fig:uzonal_d7}) and thus the longer advective timescale should lead to smaller phase curve offsets \citep{Cowan:2011,Zhang:2016}, as opposed to our results. Second, the larger radiative heating rates on the dayside of models including hydrogen dissociation (Figure \ref{fig:heatingrate}) imply that radiation more strongly forces the temperature field to the equilibrium state, which again should imply smaller phase curve offsets in models with hydrogen dissociation. A plausible explanation for our results is that the total heat budget carried by the eastward equatorial jet that is characterized by the day-night difference of $\bar{c_p}T+\mathcal{L}_h q$ is greatly enhanced by hydrogen dissociation, and this effect can overcome the two effects mentioned above and  give rise even more efficient heat transport to the eastward of the substellar point.

\subsubsection{Varying rotation}

\begin{figure}
\centering
\includegraphics[width=0.45\textwidth]{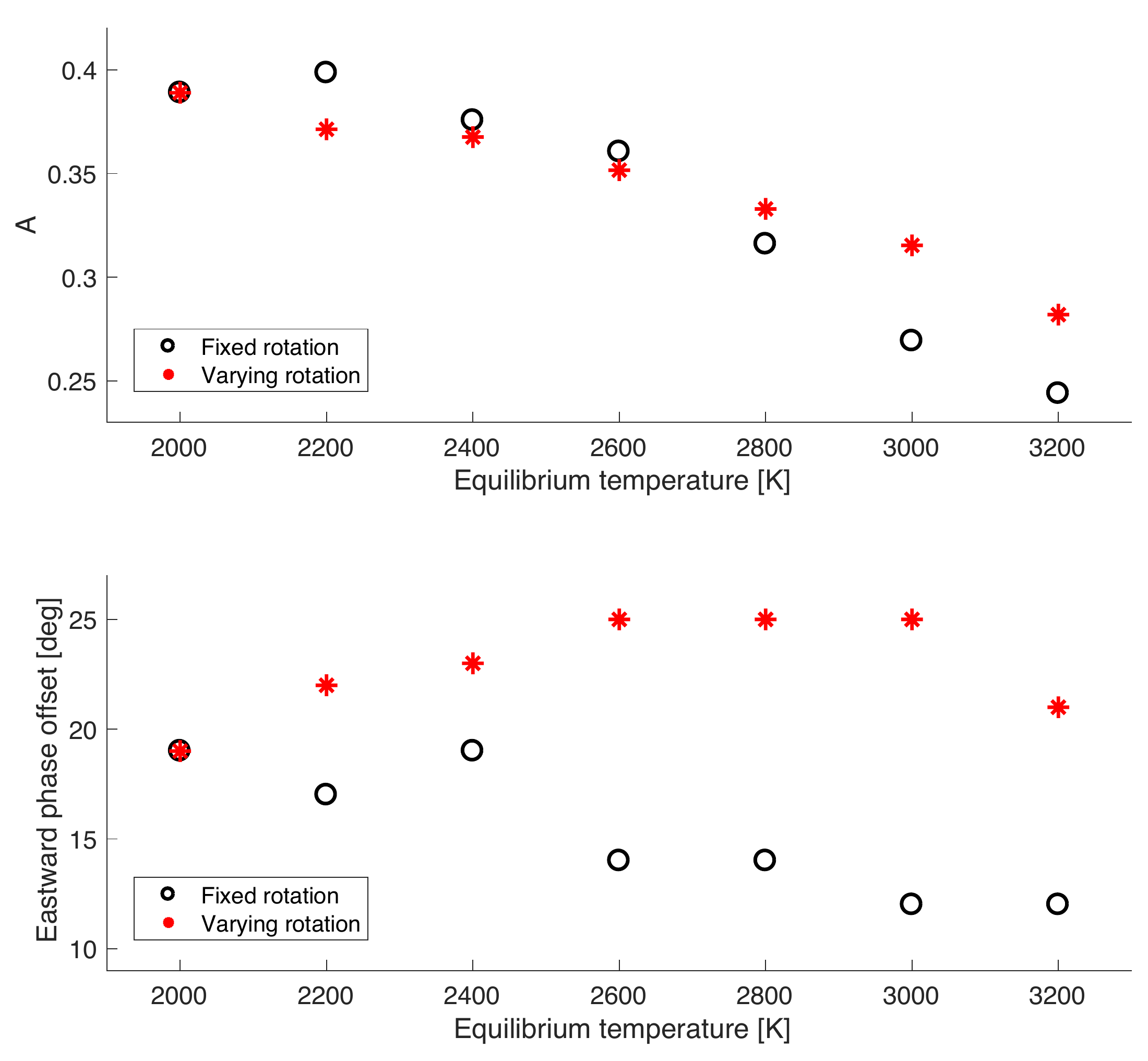}
\caption{Phase curve amplitude (top) and offset (bottom) from simulations with fixed rotation (black points) and decreasing rotation period with increasing equilibrium temperature (red points). We find that the phase curve amplitude in our simulations with varying rotation is smaller for cool equilibrium temperatures and higher for hot equilibrium temperatures than in our simulations with fixed rotation. The phase offset is always larger in our simulations with varying rotation, due to the stronger eastward equatorial jet in our simulations with varying rotation relative to our simulations with fixed rotation.}
  \label{fig:fastrot_phasecurve}
\end{figure}

\Fig{fig:fastrot_phasecurve} compares the normalized phase curve amplitude and offset from simulations that have a fixed rotation and that include a decrease in the rotation period with increasing equilibrium temperature.  The normalized phase curve amplitude, denoted as $\mathcal{A}$, is defined as $1-(F_{\rm{min}}/F_{\rm{max}})^{1/4}$ where $F_{\rm{min}}$ and $F_{\rm{max}}$ are the minimum and maximum flux in the phase curve, respectively.  Both of these suites of simulations include weak drag and the effects of hydrogen dissociation and recombination. We find that the phase curve amplitude decreases with increasing equilibrium temperature in the ultra-hot Jupiter regime for both suites of simulations with a fixed and varying rotation period. This is a result of the increasing efficiency of day-to-night heat transport by hydrogen dissociation with increasing equilibrium temperature. However, the slope of the decrease in the phase curve amplitude is smaller for the simulations with varying rotation. At $T_{\rm{eq}}\lesssim 2800$ K, the phase curve amplitudes of models with varying rotation are slightly smaller than those with a fixed rotation due to the enhanced day-to-night heat transport by the stronger equatorial jets in rapid rotators. At $T_{\rm{eq}}\gtrsim 2800$ K, the phase curve amplitudes are larger for rapid rotators presumably because the large day-night temperature difference at off-equatorial regions due to stronger rotation contributes more to the day-night thermal difference.

\indent The phase curve offset shows opposite trends with varying equilibrium temperature in our simulations with fixed and varying rotation. For our simulations with fixed rotation, the phase offset decreases with increasing equilibrium temperature. This is because firstly, the dayside of the planet warms with increasing equilibrium temperature, which tends to reduce the radiative timescales and causes smaller hot spot offsets. Secondly, the equatorial jet is weakened with increasing equilibrium temperature for fixed-rotation models, and the increase of the advective timescale also tends to reduce the hot spot offsets. However, in our simulations with varying rotation the phase offsets are in general much larger than those with a fixed rotation, simply because when varying rotation, the equatorial jet is stronger than that with fixed rotation. For models with varying rotation, the phase curve offset first increases and then flats out with increasing equilibrium temperature, although the speed of the equatorial jet actually slightly decreases from $T_{\rm{eq}}=2200$ K to 3000 K. The increase in the heat budget $\bar{c_p}T+\mathcal{L}_h q$ with increasing equilibrium temperature is likely responsible for this behavior.  Interestingly, although the equatorial jet becomes stronger again at $T_{\rm{eq}}=3200$ K, the phase curve offset decreases. This is likely a result of the rapid change of the horizontal wave patterns at $T_{\rm{eq}}=3200$ K (see Figure \ref{fig:temp_d7_rotconsis}) -- the off-equatorial Rossby waves are much more confined toward the equator than those with $T_{\rm{eq}}=3000$ K, and the hot areas associated with the westward-shifted Rossby gyres counteract the eastward hot spot displacement at the equator, and thus reduce the phase curve offset.

\subsubsection{Time-variability}
\label{sec:variability}

\begin{figure}
\centering
\includegraphics[width=0.45\textwidth]{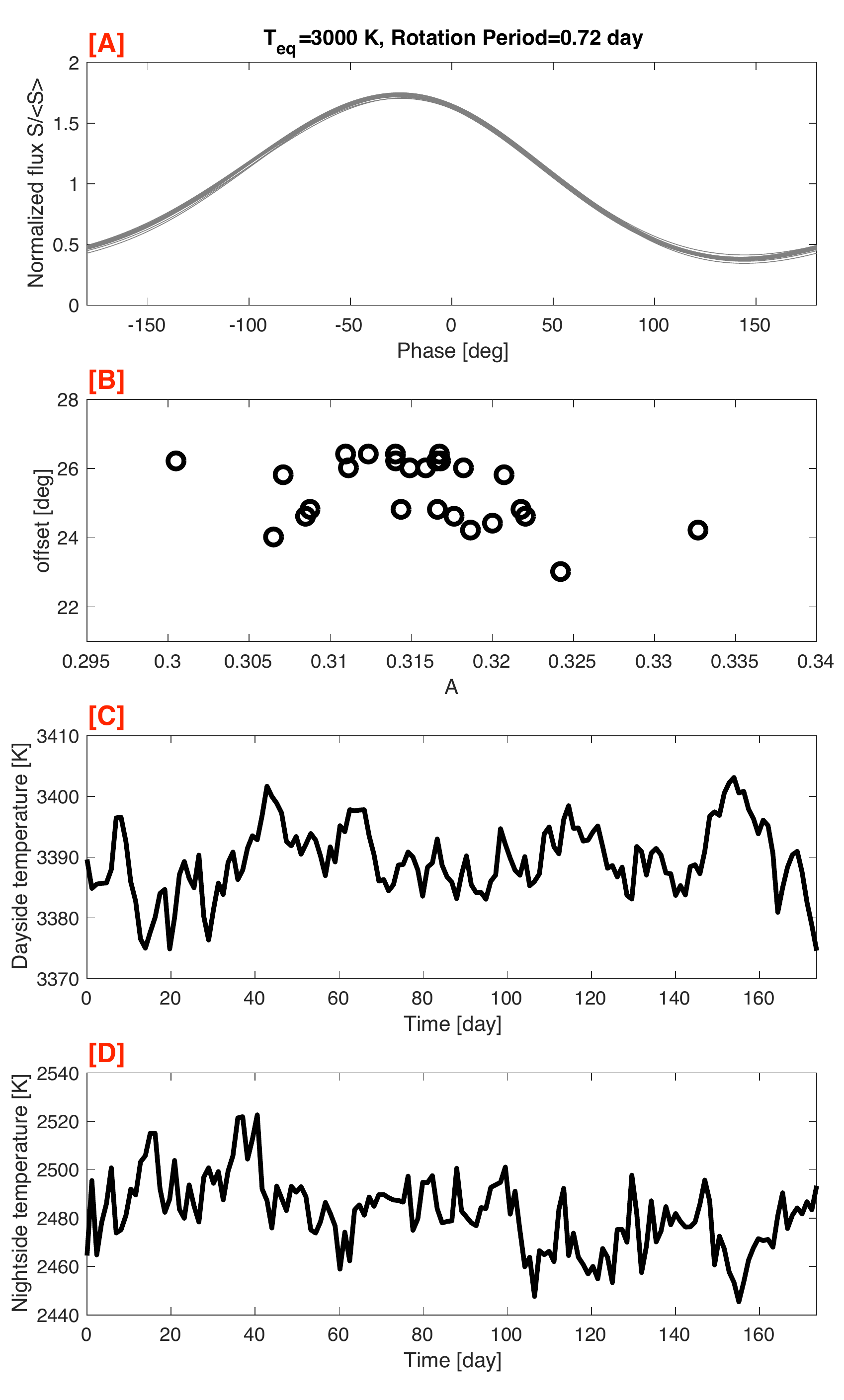}
\caption{Variability in a simulation including dissociation that has an equilibrium temperature of 3000 K, rotation period of 0.72 days, and drag timescale of $10^7 \ \mathrm{s}$. Panel [A]: Normalized phase curves over the last 500 days of simulation time. Panel [B]: phase curve offset and amplitude from this sample of phase curves. We find that over a 500-day baseline the phase curve offset varies by up to $\sim 3^\circ$ and the phase curve amplitude varies by up to $\sim 10\%$. Panels [C] and [D]: time evolution of nightside and dayside brightness temperatures over 170 days of simulation time. We find that the nightside temperature can vary by up to $\sim 80 \ \mathrm{K}$, while the dayside temperature varies by $\sim 30 \ \mathrm{K}$.} 
  \label{fig:variability}
\end{figure}

Variability due to generation and evolution of vortices and the interactions with the mean flow in rapidly rotating models ($T_{\rm{eq}}\gtrsim 3000$ K and $P_{\rm{rot}}\lesssim 0.72$ days) with weak drag can lead to time-variable phase curves and dayside and nightside thermal flux. \Fig{fig:variability} shows the variability in the phase curve, phase offset and amplitude, and dayside and nightside brightness temperatures from our simulation with an equilibrium temperature of 3000 K, weak drag, and a rotation period of 0.72 days.  Quantities in panel [A] and [B] are sampled across about 500 days of simulation time. The dayside and nightside brightness temperature are defined as $(F_{\rm{day}}/\sigma)^{1/4}$ and $(F_{\rm{night}}/\sigma)^{1/4}$, respectively, where $\sigma$ is the Stefan-Boltzmann constant, $F_{\rm{day}}$ is the phase curve flux at the secondary eclipse and $F_{\rm{night}}$ is the phase curve flux at the primary transit.  In panel [C] and [D] we show time evolution of the dayside and nightside brightness temperature over 170 days of simulation time. We find that the nightside temperature can vary by up to $\sim 80 \ \mathrm{K}$ and the dayside temperature can vary by up to $\sim 30 \ \mathrm{K}$ over the last 500 days of simulation time. The fractional variability is larger in the nightside flux than in the dayside flux. This is partly because the vortices are largely active at high latitudes on the nightside of the planet. The time evolution of the dayside and nightside brightness temperatures exhibit a chaotic nature. But a typical timescale between several days and more than ten days seems to emerge, corresponding to the typical life cycles of the off-equatorial vortices. There seem to be longer timescsles of several tens of days, which might be due to more complex interactions between the vortices and the mean flow. We defer the detailed study of such variability to future work.  

\indent Though the variability in hemisphere-averaged temperatures is relatively small, we find that the variability in observable properties is significant. We find from the same simulation as above that the phase curve amplitude varies by up to $\sim 10\%$, and the phase curve offset varies by up to $3^\circ$. This variation in the phase curve amplitude could be detectable, as upper limits on secondary eclipse variability of a few percent have been placed on HD 189733b and HD 209458b from previous \textit{Spitzer} observations \citep{Agol:2010,Kilpatrick:2019aa}, and secondary eclipse variability has been found in the optical wavelength \textit{Kepler} light curves of Kepler-76b \citep{Jackson:2019aa}. If such time-variability caused by large-scale instability occurs in the atmospheres of ultra-hot Jupiters, its effect on the infrared phase curve amplitude may be detectable with current instrumentation.

\section{Comparison to observations}
\label{sec:comptoobs}

\subsection{Day-night temperature contrast and phase curve offset}

\begin{figure*}
\centering
\includegraphics[width=1\textwidth]{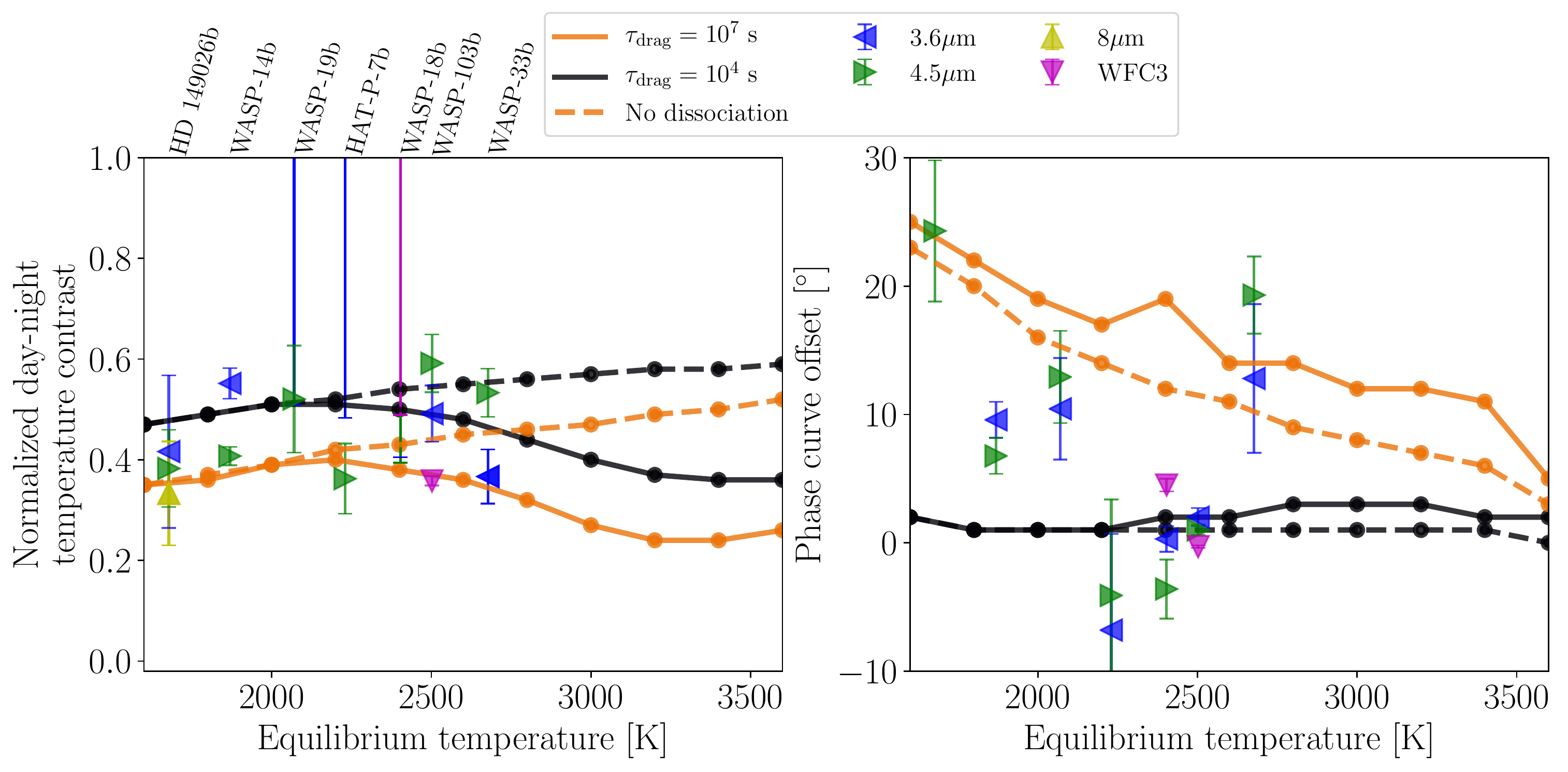}
\caption{GCM predictions for the day-night brightness temperature contrast and phase curve offset compared to observations. Left hand panel: GCM predictions for the day-night brightness temperature contrast (points with lines connecting them) compared to \textit{Spitzer} and \textit{HST/WFC3} observations. The solid lines are GCM results including the effects of hydrogen dissociation and recombination, while the dashed lines do not include these effects. The GCM experiments use a fixed rotation period of 2.43 days with varying equilibrium temperature. The orange line is for a case with weak drag ($\tau_\mathrm{drag} = 10^7 \ \mathrm{s}$), while the black line is for a case with strong drag ($\tau_\mathrm{drag} = 10^4 \ \mathrm{s}$). Right hand panel: GCM predictions for the phase curve offset compared to observations. Data are shown for the same planets that are labeled on the left hand panel. We find that the GCM results generally bracket the observed range of day-night brightness temperature contrast and phase curve offsets. Note that we cannot explain the westward offset of $-32.2^\circ$ for HD 149026b at $3.6 \ \mu\mathrm{m}$ \citep{Zhang:2017a}, which lies below the y-axis scale. Additionally, we find that the day-night brightness temperature contrast decreases with increasing equilibrium temperature in the ultra-hot Jupiter regime, as expected from analytic theory \citep{Bell:2018aa,Komacek:2018aa}.}
  \label{fig:ampoffsetdata}
\end{figure*}

\indent \Fig{fig:ampoffsetdata} compares the normalized  phase curve amplitude (which is equivalent to the normalized day-night temperature contrast)  and phase curve offset predicted by our suite of GCM experiments with \textit{Spitzer} and \textit{Hubble} phase curve observations\footnote{Observations are from \cite{Knutson_2007,Knutson:2009a,Knutson:2009,Nymeyer:2011,Cowan:2012,Crossfield:2012,Knutson:2012,Maxted:2013,Stevenson:2014,Zellem:2014,Wong:2015a,Wong:2015,Stevenson2016,Zhang:2017a,Kreidberg:2018aa}, and \cite{Arcangeli:2019aa}.}. 
The normalized day-night temperature contrast from our GCM experiments including hydrogen dissociation and recombination decreases with increasing equilibrium temperature in the ultra-hot Jupiter temperature regime, as expected from \cite{Bell:2018aa} and \cite{Komacek:2018aa}. Meanwhile, the day-night temperature contrast from our simulations that do not include hydrogen dissociation and recombination have an increasing day-night temperature contrast with increasing equilibrium temperature, as expected from previous modeling work and observations \citep{Cowan_2011,Perez-Becker:2013fv,Komacek:2015,Schwartz:2015,Komacek:2017,Schwartz:2017aa}. The phase curve offset from our GCM experiments with weak drag ($\tau_\mathrm{drag} = 10^7 \ \mathrm{s}$) decrease with increasing equilibrium temperature due to the decreasing radiative damping timescale with increasing equilibrium temperature \citep{Zhang:2016}. As has been explained for Figure \ref{fig:phasecurves_d7}, the phase curve offsets from GCM experiments that include hydrogen dissociation and recombination are overall slightly larger than those that do not include hydrogen dissociation and recombination. This is because the heat budget is enhanced by including hydrogen dissociation.

\indent In general, we find from \Fig{fig:ampoffsetdata} that our GCM results with varying drag timescale roughly bracket the observed day-night brightness temperature contrasts. We find that hydrogen dissociation and recombination is required to explain the low day-night temperature contrasts observed for WASP-103b with \textit{HST/WFC3} and WASP-33b with \textit{Spitzer} Channel 1. Our models that include hydrogen dissociation and recombination have significantly decreased day-night temperature contrasts for $T_\mathrm{eq} \gtrsim 3000 \ \mathrm{K}$. Additionally, the difference in predicted day-night temperature contrast between our models with and without hydrogen dissociation and recombination is largest for the hottest planets considered in our modeling suite. As a result, phase curves of hot Jupiters that receive greater incident stellar flux than WASP-103b and WASP-33b will provide a powerful test of how hydrogen dissociation and recombination shape the atmospheric circulation of ultra-hot Jupiters.

\indent All simulations including hydrogen dissociation and recombination have higher phase offsets than simulations not including hydrogen dissociation and recombination. We also find that the phase offsets calculated from our suite of GCM experiments generally bracket observations. For our simulations with weak drag ($\tau_\mathrm{drag} = 10^7 \ \mathrm{s}$), the phase curve offset decreases with increasing equilibrium temperature for simulations both with and without hydrogen dissociation and recombination. For simulations with strong drag ($\tau_\mathrm{drag} = 10^4 \ \mathrm{s}$), the phase curve offsets are always close to zero, with a slight increase at high equilibrium temperatures when including hydrogen dissociation and recombination due to the cooler dayside. Although small, the non-zero phase offsets in the strong-drag models including hydrogen dissociation are interesting. Zonal jets are entirely suppressed in this regime, and thus are not responsible for the phase offsets. A plausible explanation is that the effects of hydrogen dissociation and recombination cause a faster phase speed for the eastward Kelvin waves compared to those without hydrogen dissociation, such that fast Kelvin wave propagation could still cause eastward hot spot displacement under strong drag. There may be two reasonings for the faster Kelvin wave phase speed. First, an increase in the fraction of atomic hydrogen increases the scale height, which increases the gravity wave speed. The second reasoning is related to the fact that  the fractional atomic hydrogen increases with decreasing pressure. Suppose that a wave packet ascends and experiences cooling due to the increasing fraction of atomic hydrogen, and it gains extra downward buoyancy compared to ``dry" gravity waves. This is as if the stratification is enhanced. This is an interesting prospect for future study of diabatic waves because moist, condensible processes on earth's troposphere usually tend to slow down the phase speed of gravity waves (e.g., \citealp{kiladis2009}).  A theoretical framework is needed to establish a better understanding of waves in atmospheres of ultra-hot Jupiters. 

Note that we only consider planets with zero obliquity, while non-zero obliquities could explain observed westward phase offsets \citep{L.-Dang-et-al.:2018aa,Adams:2019aa}. From our simulations with fixed rotation, we do not find a change in sign from decreasing to increasing phase curve offsets with increasing equilibrium temperature, as was suggested by \cite{Zhang:2017a}. However, when decreasing the rotation period with increasing equilibrium temperature (\Fig{fig:fastrot_phasecurve}), we do find that hotter planets with shorter rotation periods have larger phase curve offsets. 
\subsection{Dayside and nightside temperature}

\begin{figure*}
\centering
\includegraphics[width=1\textwidth]{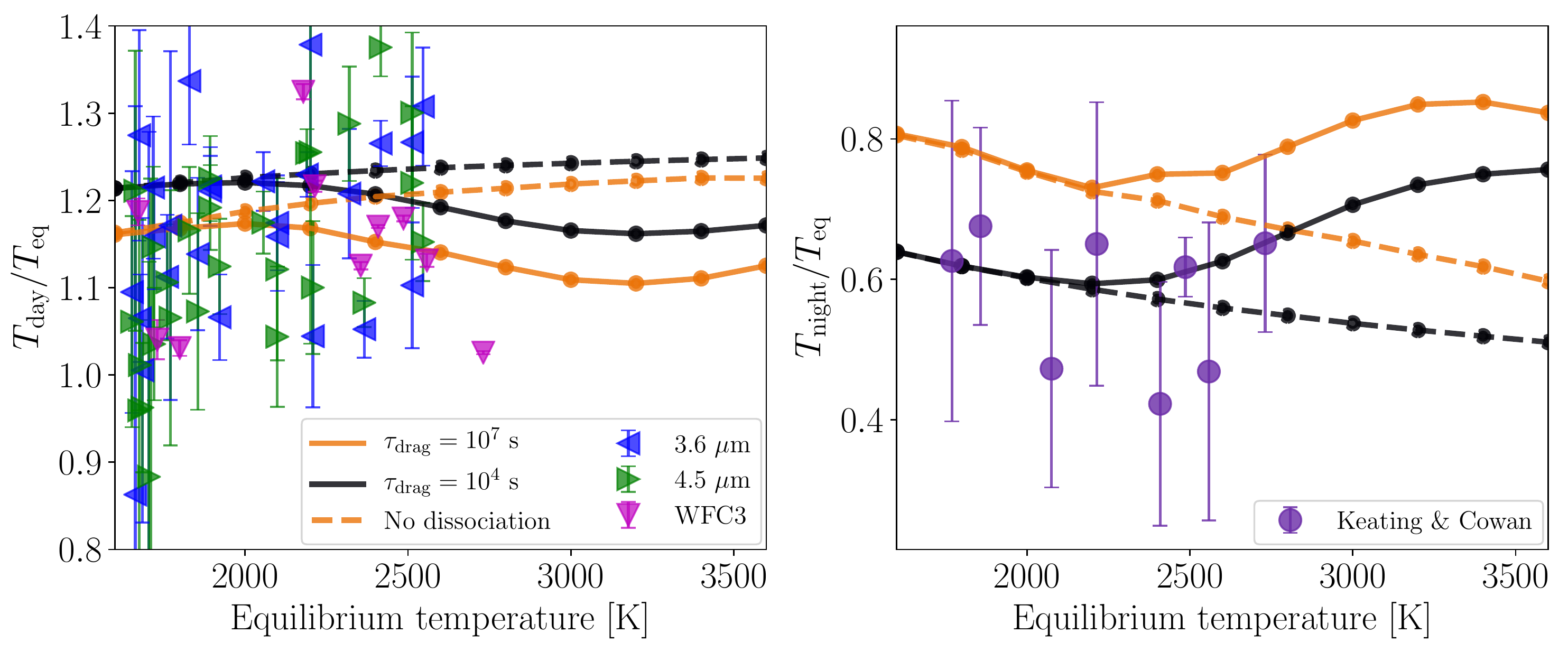}
\caption{Dayside and nightside brightness temperatures normalized to the equilibrium temperature, from our GCM experiments (points with lines) compared to observations (points). Left hand panel: dayside temperature vs. equilibrium temperature. Solid lines show GCM results including the effects of hydrogen dissociation and recombination, dashed lines show results that do not include these effects. The GCM experiments use a fixed rotation period of 2.43 days. Blue and green points are from are from \textit{Spitzer} Channels 1 and 2, and magenta points are from \textit{Hubble/WFC3}. Right panel: nightside temperature normalized to the equilibrium temperature vs. equilibrium temperature. Points show the nightside temperatures derived from the analysis of \cite{Keating:2018aa}. We find that strong drag is needed to explain the low observed nightside temperatures within the context of our GCM results.}
\label{fig:tdaytnight}
\end{figure*}

\indent \Fig{fig:tdaytnight} compares the dayside and nightside brightness temperatures (normalized to the equilibrium temperature) calculated from our GCM experiments with those observed from phase curves and at secondary eclipse\footnote{\textit{Spitzer} secondary eclipse data are from \cite{Garhart:2019aa}, while \textit{Hubble/WFC3} secondary eclipse data are from \cite{Parmentier:2018aa}, which compiled the secondary eclipse observations of \cite{Crouzet:2014aa,Kriedberg:2014,Ranjan:2014aa,Stevenson:2014aa,Wilkins:2014aa,Haynes:2015,Line:2016aa,Beatty:2017aa,Evans:2017aa,Arcangeli:2018aa,Kreidberg:2018aa,Mansfield:2018aa}, and \cite{Nikolov:2018aa}. Nightside temperatures are from the phase curve inversions of \cite{Keating:2018aa}. Note that many of the planets observed have either no detected nightside or negative brightness on the nightside, so \cite{Keating:2018aa} refit the data to find the nightside brightness temperature.}. Our GCM experiments without the effects of hydrogen dissociation and recombination have a ratio of the dayside temperature to the equilibrium temperature ($T_\mathrm{day}/T_\mathrm{eq}$) that always increases with increasing equilibrium temperature. However, our simulations including hydrogen dissociation and recombination have a decreasing $T_\mathrm{day}/T_\mathrm{eq}$ from $2000 \ \mathrm{K} \le T_\mathrm{eq} \le 3200 \ \mathrm{K}$. We can only explain \textit{HST/WFC3} dayside brightness temperatures of some ultra-hot Jupiters with our GCM experiments that include the effects of hydrogen dissociation and recombination. For some that have higher $T_\mathrm{day}/T_\mathrm{eq}$  than prediected by our models, the dayside atmosphere could have a thermal inversion, with the observations, especially the \textit{Spitzer} 4.5 $\mu$m observations, probing the hot, low-pressure layers. Future secondary eclipse observations of planets with $T_\mathrm{eq} \gg 2500 \ \mathrm{K}$ could help determine the impact of hydrogen dissociation and recombination on dayside temperatures of ultra-hot Jupiters. 

\indent From \Fig{fig:tdaytnight}, we find that increasing the drag strength leads to hotter daysides and cooler nightsides in our simulations. We require strong drag to explain the nightside temperatures of hot Jupiters from our simulations. Alternatively, nightside cloud decks have been proposed to explain the low observed nightside brightness temperatures of hot Jupiters \citep{Kataria:2014,Beatty:2018aa,Mendonca:2018,Keating:2018aa}. Though we do not include clouds in our GCM experiments, the nightside temperatures we calculate for planets with $T_\mathrm{eq} \lesssim 3000 \ \mathrm{K}$ are cool enough to allow for condensate (equilibrium) cloud formation \citep{Parmentier:2015,Wakeford:2017,Roman:2019aa,Helling:2019aa}, which may greatly reduce the outgoing infrared flux on the nightside \citep{Roman:2017aa,Mendonca:2018}.

\section{Discussion}
\label{sec:disc} 
\indent Our GCM results find similar qualitative trends as previous theoretical work that aimed to understand how hydrogen dissociation and recombination affect the climate of hot Jupiter atmospheres. Similarly to \cite{Bell:2018aa} and \cite{Komacek:2018aa}, we find that hydrogen dissociation and recombination act to reduce the day-night temperature contrast and resulting phase curve amplitudes of ultra-hot Jupiters (see \Fig{fig:ampoffsetdata}). As predicted by \cite{Komacek:2018aa}, we find that hydrogen dissociation and recombination cause a reduction in the speed of winds in ultra-hot Jupiter atmospheres when holding the rotation period fixed (\Fig{fig:uzonal_d7}). Additionally, we find from our GCM experiments that hydrogen dissociation and recombination act to increase the nightside temperatures of ultra-hot Jupiters (\Fig{fig:tdaytnight}) and increase the phase curve offset (\Fig{fig:ampoffsetdata}), similar to that in \cite{Bell:2018aa}.      \\ 
\indent In \Sec{sec:comptoobs}, we compared our GCM results to observations, finding that we could largely explain the range of observed dayside-to-nightside temperature contrasts and phase curve offsets with our GCM experiments including hydrogen dissociation and recombination. Our GCM results show the largest difference with and without hydrogen dissociation and recombination for equilibrium temperatures larger than 3000 K, a regime which has only one observed phase curve to date. \cite{Mansfield:2019aa} recently obtained a \textit{Spitzer} phase curve of KELT-9b, which has an equilibrium temperature of 4050 K \citep{Gaudi:2017aa}. \cite{Mansfield:2019aa} found that KELT-9b has a relatively cool dayside with $T_\mathrm{day}/T_\mathrm{eq} = 1.11 \pm 0.01$ and a small normalized day-night temperature contrast of $0.436 \pm 0.020$. Referring back to \Fig{fig:ampoffsetdata}, the normalized day-night temperature contrast of KELT-9b at $4.5 \mu\mathrm{m}$ is smaller than those of WASP-103b and WASP-33b, even though both planets have equilibrium temperatures over $1000 \ \mathrm{K}$ cooler than KELT-9b. We require hydrogen dissociation and recombination to explain the observed dayside-nightside temperature contrast of KELT-9b with our GCM (see \citealp{Mansfield:2019aa}). Future phase curve observations of planets with equilibrium temperatures between that of WASP-33b and KELT-9b will help determine the impacts of hydrogen dissociation and recombination on the atmospheric circulation of ultra-hot Jupiters. 

\cite{caldas2019} showed that the differences in temperature and chemistry across the limb could potentially have significant effects on transmission spectra which would probably not be easily captured by 1D models. We find from our models that in the upper atmosphere at pressures close to 1 mbar, the atomic hydrogen fraction is normally larger than that in the mean near-IR photosphere. Thus, the normalized day-night temperature difference at low pressures is smaller than that near the photosphere due to an increased role of hydrogen dissociation and recombination heat transport. This indicates that compared to models without hydrogen dissociation and recombination, including the dissociation and recombination heat transport would increase the temperature near the terminator. As a result, the chemistry, such as the dissociated fraction of water and the abundance of H$^-$, is expected to significantly alter transmission spectra. Besides that, as we have shown in the main text, at very low pressure the circulation tends to be dominated by day-night flow and the westward flow can be the dominant component. This change in the character of the circulation would also influence the transmission spectra.  Here we do not discuss this effect quantitatively based on our modeling results.  This is because we are cautious that our semi-grey radiative transfer overestimates the temperature and radiative timescale at low pressures, both of which are crucial to calculate precise thermal structures at low pressure and thus make predictions for transmission spectra. This is an intrinsic deficit of the semi-grey approximation of the radiative transfer. We defer detailed studies of the effects of hydrogen dissociation and recombination heat transport on transmission spectroscopy to future models that include realistic radiative transfer and chemistry. 

\indent Our GCM experiments were simplified in order to study a wide parameter space of possible ultra-hot Jupiters. As a result, we did not include non-grey radiative transfer, cloud formation and cloud radiative effects, and the effects of varying opacity with equilibrium temperature. We caution about the simplicity of our radiative transfer scheme. For example, ultra-hot Jupiters likely have thermal inversions due to absorption by H$^{-}$ opacity, which is not included here. Its role in the atmospheric circulation includes several aspects. First, the additional absorption due to H$^{-}$ causes the stratification to be enhanced, resulting in an increased  deformation radius. As a result, the meridional extent of the equatorial jet could be widened. Second, the day-night equilibrium temperature difference at the near-IR photosphere increases due to the dayside thermal inversion, and the radiative timescale decreases due to stronger absorption of stellar insolation, both of which tend to increase the day-night temperature difference. Third, however, the day-night difference of atomic hydrogen fraction increases, which helps to reduce the day-night temperature contrast. We defer an analysis of how these combined effects of H$^{-}$ opacity alter circulation to future work. Additionally, we did not include the detailed effects of magnetohydrodynamics, which have been shown to significantly affect the phase curve offsets of ultra-hot Jupiters \citep{Rogers:2017}. Though we considered varying equilibrium temperature, drag strength, and rotation period, we did not include the effect of varying gravity or atmospheric composition. Future work could explore how including more realistic radiative transfer changes how hydrogen dissociation and recombination affect the day-night temperature contrast and phase curve offset of ultra-hot Jupiters. By coupling non-grey radiative transfer with our implementation for the effects of hydrogen dissociation and recombination, one could make predictions for how hydrogen dissociation and recombination affect spectro-photometric phase curves of ultra-hot Jupiters. 

Finally, because thermal dissociation of molecular hydrogen is likely inevitable in atmospheres of ultra-hot Jupiters, their substantial effect in reducing the dayside temperature and increasing the nightside temperature is robust. By comparing phase-curve and secondary-eclipse observations to detailed circulation modelling, this could help understand other important processes in the atmospheres of ultra-hot Jupiter in an indirect way, for instance, constraining nightside clouds, dayside thermal inversions or the effects of magnetic fields.

\section{Conclusions}
\label{sec:conc}
\indent In this work, we incorporated the effects of hydrogen dissociation and recombination into a GCM for ultra-hot Jupiter atmospheres. Using this GCM, we simulated a large suite of ultra-hot Jupiters with varying equilibrium temperature, rotation period, and frictional drag strength, both including and ignoring the effects of hydrogen dissociation and recombination. We find that including hydrogen dissociation and recombination greatly affects the atmospheric circulation of ultra-hot Jupiters, cooling their daysides, warming their nightsides, and shaping their equatorial flow. We summarize the main conclusions from our work below.
\begin{enumerate}
\item We find using numerical circulation modeling that hydrogen dissociation and recombination greatly reduces the dayside-to-nightside temperature contrasts of ultra-hot Jupiters. Two effects combine to reduce the day-to-night temperature contrast. First, energy input is required to dissociate molecular hydrogen on the dayside of the planet, cooling the dayside. Second, atomic hydrogen recombines into molecular form in cooler regions, releasing heat which warms the nightside. Our GCM simulations confirm the theoretical predictions of \cite{Bell:2018aa} and \cite{Komacek:2018aa} that hydrogen dissociation and recombination reduce the day-night temperature contrasts of hot Jupiters.
\item  We find that when the planetary rotation is fixed at a canonical hot-Jupiter rotation, the strength of the equatorial jet decreases with increasing temperature for models including hydrogen dissociation in the ultra-hot Jupiter regime. The inclusion of hydrogen dissociation and recombination reduces the day-night temperature contrast, which results in a smaller wind speed. The decrease in wind speed  presumably leads to reduced wave-mean-flow interactions with increasing temperature, responsible for the weaker equatorial jets. When the planetary rotation period decreases self-consistently with increasing equilibrium temperature,  the equatorial superrotating jets are robust over different equilibrium temperature, in contrast to results with a fixed rotation period.  The strength and meridional extent of the equatorial superrotating jet are not strongly affected by the increasing cooling and heating due to hydrogen dissociation and recombination. A theoretical framework is needed to better understand wave dynamics and wave-mean-flow interactions in atmospheres of ultra-hot Jupiters.
\item The effects of hydrogen dissociation and recombination strongly affect full-phase light curves of hot Jupiters. With hydrogen dissociation and recombination, ultra-hot Jupiters have significantly reduced phase curve amplitudes and larger phase curve offsets than in models without their effects. We find that the phase curve amplitude decreases with increasing temperature in the ultra-hot Jupiter regime, due to the strong increase of the fraction of atomic hydrogen with temperature. This could provide an explanation for the relatively small phase curve amplitudes of observed ultra-hot Jupiters, including WASP-103b, WASP-33b, and KELT-9b.
\end{enumerate}

\acknowledgements
We thank Taylor Bell for helpful discussions about the physics of hydrogen dissociation and recombination. We thank Jacob Bean and Megan Mansfield for helpful comments on an early draft of this manuscript. We thank Shang-Min Tsai for helpful discussions. We thank the referee for constructive comments. This work benefited from the Exoplanet Summer Program in the Other Worlds Laboratory (OWL) at the University of California, Santa Cruz, a program funded by the Heising-Simons Foundation. Our work was completed with resources provided by Lunar and Planetry Laboratory, University of Arizona, the Department of Physics, University of Oxford, and the University of Chicago Research Computing Center. X. Tan acknowledges support from the European community through the ERC advanced grant EXOCONDENSE (PI: R.T. Pierrehumbert). T.D.K. acknowledges funding from the 51 Pegasi b Fellowship in Planetary Astronomy sponsored by the Heising-Simons Foundation.
\if\bibinc n
\bibliography{References_all}
\fi

\if\bibinc y

\fi

\end{document}